\documentclass[aps]{revtex4}

\usepackage[utf8]{inputenc} 
\usepackage{amsmath}
\usepackage{amssymb}
\usepackage{mathrsfs}
\usepackage{bm}
\usepackage{color}
\usepackage{braket}
\usepackage{subfig}
\usepackage{graphicx}

\DeclareGraphicsExtensions{.pdf,.jpg}

\newcommand{\tr}{\text{Tr}}
\renewcommand{\bf}[1]{\textbf{#1}}

\begin{document}
\author{Syed Tahir Amin$^{1,3,4}$, Bruno Mera$^{1,3,4}$, Chrysoula Vlachou$^{2,3,4}$, Nikola Paunkovi\'c$^{2,3,4}$, and V\'{\i}tor R. Vieira$^{1,4}$}
\affiliation{$^1$  Departamento de F\'{\i}sica, Instituto Superior
T\'ecnico, Universidade de Lisboa, Av. Rovisco Pais, 1049-001 Lisboa, Portugal}
 \affiliation{$^2$ Departamento de Matemática, Instituto Superior Técnico, Universidade de Lisboa, Av. Rovisco Pais, 1049-001 Lisboa, Portugal}
\affiliation{ $^3$ Instituto de Telecomunica\c{c}\~oes, 1049-001 Lisbon, Portugal}
\affiliation{$^4$ CeFEMA, Instituto Superior
T\'ecnico, Universidade de Lisboa, Av. Rovisco Pais, 1049-001 Lisboa, Portugal}

\title{Fidelity and Uhlmann connection analysis of topological phase transitions in two dimensions}

\begin{abstract}
We study the behaviour of the fidelity and the Uhlmann connection in two-dimensional systems of free fermions that exhibit non-trivial topological behavior. In particular, we use the fidelity and a quantity closely related to the Uhlmann factor in order to detect phase transitions at zero and finite temperature for topological insulators and superconductors. We show that at zero temperature both quantities predict quantum phase transitions: a sudden drop of fidelity indicates an abrupt change of the spectrum of the state, while the behavior of the Uhlmann connection signals equally rapid change in its eigenbasis. At finite temperature, the topological features are gradually smeared out, indicating the absence of finite-temperature phase transitions, which we further confirm by performing a detailed analysis of the edge states. Moreover, we performed both analytical and numerical analysis of the fidelity susceptibility in the thermodynamic limit, providing an explicit quantitative criterion for the existence of phase transitions. The critical behaviour at zero temperature is further analysed through the numerical computation of critical exponents.
\end{abstract}

\maketitle

\section{Introduction}

Topological phases of matter have received much attention in recent years due to their many potential applications in the emerging field of quantum technologies. Topological phases (TPs), in contrast to the Landau theory of phase transitions~\cite{lan:37}, are not described in terms of a local order parameter having a non trivial value in the symmetry broken phase. Instead, they are described by global topological invariants~\cite{tknn:82,ber:84,ando:13}, robust against continuous perturbations of the system preserving the symmetry and the spectral gap -- no breaking of the symmetry occurs. These topological invariants, like Chern numbers~\cite{tknn:82}, become observable through the response functions of the system coupled to an external field~\cite{tknn:82,mer:17}. When terminating the system to the vacuum, there appear gapless modes which are exponentially localized at the boundary of the system, as predicted by the bulk-to-boundary principle~\cite{ryu:hat:02}. On the other hand, it is impossible to deform a system from a topological phase with a given value of a topological invariant to the one where this invariant takes a different value without closing the gap. In this case, we say that a topological phase transition occurred, a particular example of a quantum phase transition~\cite{sac:07}. From the quantum information theory point of view, there are two main approaches to the problem of quantum phase transitions. The first one is to study entanglement properties of the system~\cite{ost:ami:fal:faz:02,vid:lat:ric:kit:03,su:son:gu:06,oli:sac:14}. The second approach is to study various distinguishability measures between quantum states, in particular the quantum fidelity~\cite{uhl:76,alb:83,zan:qua:wan:sun:06,zan:pau:06,pau:sac:nog:vie:dug:08,sac:pau:vie:11,oli:sac:14}, which we will consider here. When a system undergoes a phase transition, its ground state changes substantially and the quantum fidelity signals this change by a sudden drop in its value~\cite{zan:qua:wan:sun:06,zan:pau:06}. In this work, we extend the fidelity and Uhlmann connection analysis performed in~\cite{mer:vla:pau:vie:17,mer:vla:pau:vie:17:qw} for the case of chiral topological phases in 1D to 2D fermion topological phases, namely topological superconductors (TSCs) and insulators (TIs) classified by the 1st Chern number. 

Our paper is organized as follows. In Section II we recall the notions of fidelity and Uhlmann connection and how their study can be used to probe phase transitions. In Section III we present the results of the fidelity and Uhlmann connection analysis for 2D TSCs and TIs. In Section IV, we analyse the non-commutativity of the thermodynamic and zero temperature limits of the fidelity susceptibility and prove the absence of finite temperature phase transitions in the class of models considered. Moreover, we analyse the critical behaviour at zero temperature by looking at critical exponents. In Section V we consider the system in a cylinder and study the edge states. Finally, in Section VI, we present our conclusions.

\section{The fidelity and the Uhlmann connection in the study of phase transitions}
In this section, we present the main concepts of the fidelity and the Uhlmann connection analysis of phase transitions. In particular, we will show the relationship between the fidelity and the quantity $\Delta$  associated to the Uhlmann factor, which we will in turn use to study phase transitions. 

The general expression for the fidelity between two mixed states $\rho$ and $\rho'$ is
\begin{align}
F(\rho,\rho') & =\tr\sqrt{\sqrt{\rho}\rho'\sqrt{\rho}}.
\end{align}
The set of mixed states is convex, but not linear in general,  i.e., for any two mixed states $\rho$ and $\rho'$, and scalars $\lambda$ and $\lambda'$, the linear combination $\lambda\rho+\lambda'\rho'$ is not necessarily a mixed state. Nevertheless, a convex combination of $\rho$ and $\rho'$ is a mixed state, i.e., for $\lambda,\lambda'\geq 0$ and $\lambda+\lambda'=1$, the linear combination $\lambda\rho+\lambda'\rho'$ belongs to the set of mixed states. This feature of non-linearity imposes significant restrictions when it comes to performing a geometric study. On the other hand, we do not have this issue in the case of pure states, since a pure state $\rho=\ket{\psi}\bra{\psi}$ can be treated as a projection on the subspace spanned by the state $\ket{\psi}$, therefore inheriting the geometric properties of the Hilbert space. Notice the $U(1)$-gauge freedom: $\ket{\psi}$ and $e^{i\phi}\ket{\psi}$ correspond to the same state $\rho=\ket{\psi}\bra{\psi}$.
 
To overcome such restrictions, one can consider the concept of the purification of a mixed state, that is, any mixed state in a certain Hilbert space can be seen as the reduced state of a pure state in a different (larger) Hilbert space. Similarly, for each mixed state $\rho$ in a Hilbert space $H$, we can consider the corresponding Hilbert-Schmidt space $H_w=H\otimes H^*$, where $H^*$ is the dual of $H$, which contains the purifications $w$ (in this case they are called amplitudes) such that $\rho=ww^\dagger$. Notice that there is a $U(n)$-gauge freedom in the choice of the amplitude, analogously to the $U(1)$-gauge freedom in the case of pure states: $w$ and $wU$ with $U$ being unitary correspond to the same state $\rho=ww^\dagger$. In what follows, we will describe how a particular choice of the amplitude reveals the relationship between the fidelity and the Uhlmann connection. Two amplitudes $w_{1}$ and $w_{2}$, such that $\rho_{1}=w_1 w_1^\dagger$ and $\rho_{2}=w_2 w_2^\dagger$, are said to be parallel in the Uhlmann sense if they minimize the distance induced by the Hilbert-Schmidt inner product $\langle w_{2},w_{1}\rangle=\tr(w_{2}^{\dagger}w_{1})$: 
\begin{equation*}
||w_2-w_1||^2=\tr[(w_2-w_1)^\dagger (w_2-w_1)]=2(1-\text{Re}\langle w_2,w_1\rangle).	
\end{equation*}

Minimizing $||w_2-w_1||^2$ is equivalent to maximazing $\text{Re}\langle w_{2},w_{1}\rangle$: 
\begin{align*}
\text{Re}\langle w_{2},w_{1}\rangle\leq|\langle w_{2},w_{1}\rangle| & =|\tr(w_{2}^{\dagger}w_1)|.
\end{align*}
Considering the polar decompositions $w_{i}=\sqrt{\rho_{i}}U_{i},i\in\{1,2\}$, where the $U_i$'s are unitary matrices, the above inequality becomes
\begin{align*}
\text{Re}\langle w_{2},w_{1}\rangle \leq 
|\tr(U_{2}^{\dagger}\sqrt{\rho_{2}}\sqrt{\rho_{1}}U_{1})|.
\end{align*}
Using the polar decomposition $\sqrt{\rho_{2}}\sqrt{\rho_{1}}=|\sqrt{\rho_{2}}\sqrt{\rho_{1}}|V$, with $V$ unitary, and the cyclic property of the trace, we obtain
\begin{align*}
\text{Re}\langle w_{2},w_{1}\rangle \leq 
|\tr(|\sqrt{\rho_{2}}\sqrt{\rho_{1}} & |VU_{1}U_{2}^{\dagger})|,
\end{align*}
and the Cauchy-Schwarz inequality implies 
\begin{align*}
\text{Re}\langle w_{2},w_{1}\rangle \leq 
\tr|\sqrt{\rho_{2}}\sqrt{\rho_{1}}|,
\end{align*}
with the equality holding for $V U_1 U_2^\dagger=I$, where $I$ is the identity~\footnote{Let us stress, that the Cauchy-Schwarz inequality, in general, reads $|\langle A,B\rangle |\leq ||A||\cdot ||B||$, and here we have considered $A=|\sqrt{\rho_1}\sqrt{\rho_2}|^{1/2}$ and $B=|\sqrt{\rho_1}\sqrt{\rho_2}|^{1/2} V U_1 U_2^{\dagger}$ to obtain the upper bound of $\text{Re} \langle w_1,w_2\rangle$. Notice that if we had considered $A= |\sqrt{\rho_1}\sqrt{\rho_2}|$ and $B=V U_1 U_2^{\dagger}$, then we would have had an extra multiplicative factor, the trace of the identity (i.e., the dimension of the Hilbert space).}. Finally, we can write $|\sqrt{\rho_{2}}\sqrt{\rho_{1}}|=\sqrt{(\sqrt{\rho_{2}}\sqrt{\rho_{1}})^{\dagger}(\sqrt{\rho_{2}}\sqrt{\rho_{1}})}$ and get
\begin{align*}
\text{Re}\langle w_{2},w_{1}\rangle \leq 
\tr\sqrt{(\sqrt{\rho_{1}}\rho_{2}  \sqrt{\rho_{1}})}=F(\rho_{1},\rho_{2}),
\end{align*}
which shows the relationship between the fidelity and the Uhlmann connection, as characterized by $V$, the so-called Uhlmann factor. In other words, imposing two amplitudes, $w_1$ and $w_2$, to be parallel in the Uhlmann sense is equivalent to choosing, by means of the associated Uhlmann factor, the specific gauge $VU_1U_2^{\dagger}=I$, which gives rise to the fidelity between the states $\rho_1$ and $\rho_2$.

To illustrate this relationship better, let us consider the Uhlmann parallel transport condition. Suppose that $\rho(t)$ is a closed curve of density matrices parametrized by $t \in [0,1]$. Then, given an initial state $\rho(0)$ and the corresponding amplitude $w(0)$, the Uhlmann parallel transport condition, taken for an infinitesimal time period $\delta t$, yields a unique curve in the space of amplitudes, along which $w(t)$ and $w(t+\delta t)$ are parallel, $\forall t$. In particular, the Uhlmann parallel transport operator is given as $V(t)=\mathcal{T}\exp\{-\int_0^t\mathcal{A}(d\rho/ds)ds\}$, where $\mathcal{T}$ is the time-ordering operator and $\mathcal{A}$ is the Uhlmann connection differential 1-form. The Uhlmann factor is then $V=V(t+\delta t)V^{\dagger}(t)$. The length of the curve in the space of amplitudes -- with respect to the metric induced by the Hilbert-Schmidt inner product -- is equal to the length of the corresponding curve in the space of density matrices -- with respect to the Bures metric. The respective Bures distance is given in terms of the fidelity as $d_B(\rho_1,\rho_2)=\sqrt{2(1-F(\rho_1,\rho_2))}$. 

We proceed by presenting how we can apply the above concepts in the study of phase transitions.  Consider two close points $t$ and $t+\delta t$ in the parameter space and  the respective states $\rho(t)$ and $\rho(t+\delta t)$ in the space of density matrices. If the two states belong to the same phase, they almost commute, hence $V\approx I$ and $\sqrt{\rho(t+\delta t)}\sqrt{\rho(t)}\approx|\sqrt{\rho(t+\delta t)}\sqrt{\rho(t)}|$. Moreover, since they are almost indistinguishable, $F(\rho(t),\rho(t+\delta t))\approx 1$. On the other hand, if $\rho(t)$ and $\rho(t+\delta t)$ belong to different phases, they must be significantly different, thus their fidelity must be smaller than one~\cite{zan:pau:06}. The difference between the two states can be in their spectra or their eigenbases. In the case of the latter, we also have nontrivial $V\neq I$~\cite{pau:vie:08}.

To quantify the difference between the Uhlmann factor $V$ and the identity, we will use the following quantity, as defined in~\cite{pau:vie:08}:

\begin{equation}
\Delta(\rho(t),\rho(t+\delta t)) :=  
F(\rho(t),\rho(t+\delta t))- \tr(\sqrt{\rho(t)}\sqrt{\rho(t+\delta t)}).
\label{eq:deltafid}
\end{equation} 

Taking into account that $F(\rho(t),\rho(t+\delta t))= \tr(|\sqrt{\rho(t)}\sqrt{\rho(t+\delta t)}|)$ and the polar decomposition of $\sqrt{\rho(t)}\sqrt{\rho(t+\delta t)}$, as presented above, we can write:
\begin{equation}
\Delta(\rho(t),\rho(t+\delta t)) = \tr\{|\sqrt{\rho(t+\delta t)}\sqrt{\rho(t)}|(I-V)\}.
\label{eq:deltatrace}\end{equation}
Indeed, $\Delta(\rho(t),\rho(t+\delta t))$ quantifies the difference between $V$ and $I$, as for $V=I$ it is trivially zero. 

The quantity $\Delta$, just like the fidelity, is bounded by $0$ and $1$. To show this we recall the following theorem (see for example~\cite{nie:chu:04}):
\begin{equation}
|\tr (AU)|\leq \tr|A|,
\label{eq:property}
\end{equation} 
for $U$ being unitary and $A$ any operator. Using the linearity of the trace and Equation~\eqref{eq:deltatrace}, we can write $\Delta$ as:
\begin{equation}\Delta(\rho,\rho')=\tr\{|\sqrt{\rho}\sqrt{\rho'}|\}-\tr\{|\sqrt{\rho}\sqrt{\rho'}|V\}.\nonumber
\end{equation}
From~\eqref{eq:property}, by replacing $A=|\sqrt{\rho}\sqrt{\rho'}|$ and $U=V$, we obtain:
$$|\tr\{|\sqrt{\rho}\sqrt{\rho'}|V\}|\leq\tr\{|\sqrt{\rho}\sqrt{\rho'}|\}.$$
Thus, we find the lower bound of $\Delta(\rho,\rho')$ to be:
\begin{equation}
\Delta(\rho,\rho')\geq 0.
\label{eq:lowb}
\end{equation}
For the upper bound of $\Delta(\rho,\rho')$ we will use Equation~\eqref{eq:deltafid}. Since $\tr\{\sqrt{\rho}\sqrt{\rho'}\}\geq 0$, as $\sqrt{\rho}\ \text{and}\ \sqrt{\rho'}$ are positive operators, $\Delta(\rho,\rho')$ is bounded from above by the maximum value of the fidelity:
\begin{equation}
\Delta(\rho,\rho')\leq 1.
\label{eq:upperb}
\end{equation}
Summarizing, from~\eqref{eq:lowb} and~\eqref{eq:upperb}, we get:
\begin{equation}
0\leq\Delta(\rho,\rho')\leq 1.
\end{equation}
For $\rho(t)$ and $\rho(t+\delta t)$ from the same phase, $F(\rho(t),\rho(t+\delta t))=\tr(\sqrt{\rho(t+\delta t)}\sqrt{\rho(t)})\approx 1$, therefore $\Delta(\rho(t),\rho(t+\delta t))\approx0$, while for $\rho(t)$ and $\rho(t+\delta t)$ from different phases, $F(\rho(t),\rho(t+\delta t))\neq 1$ and in the case the Uhlmann factor is also non-trivial, we have $\Delta(\rho(t),\rho(t+\delta t))\neq0$. 

In summary, the departure of fidelity from 1 and the departure of $\Delta$ from 0 are indicating the phase transition points. In the particular case of topological insulators and superconductors, the respective phase transitions at zero temperature (topological quantum phase transitions) are featured by the closing of the energy gap, which occurs for certain critical values of the parameters of the Hamiltonian. The fidelity and $\Delta$ are expected to capture the gap-closing points, as the distinguishability of the states is enhanced close to the phase transition.

Regarding possible experimental observations of the fidelity and the Uhlmann connection, note that the fidelity induces the Bures metric on the space of density operators, which is nothing but the fidelity susceptibility. For large classes of orders, such as the symmetry-breaking orders of the Landau-Ginsburg type, it has been shown that the fidelity susceptibility is precisely the dynamical susceptibility of the system~\cite{ven:zan:07,zan:gio:coz:07,pau:vie:08}. Thus, measuring the susceptibility directly reveals the fidelity-induced metric in such cases. Moreover, one can measure other directly observable quantities that are shown to be related to the fidelity and its metric, such is the case of tensor monopoles~\cite{pal:gol:18}. In connection to that, it has been shown that estimating various parameters is enhanced around the regions of criticality where the fidelity-induced metric (and the related Fisher information) exhibits singularities~\cite{zan:par:ven:80}. Thus, experiencing enhanced precision in parameter estimation is directly linked to the behaviour of the fidelity, see for example a recent work~\cite{car:spa:val:18}.

As far as the Uhlmann connection is concerned, one can directly observe the associated phase, as in~\cite{viy:riv:gas:wal:fil:mar:18}. Alternatively, one can, as in the case of the fidelity, measure other directly observable physical quantities, such as dynamical susceptibilities and conductivities, that have been shown to directly depend on the mean Uhlmann curvature and the Uhlmann number introduced in~\cite{leo:val:spa:car:18}.

\section{Fidelity and $\Delta$ analysis of 2D topological Systems}
In this section we present our quantitative results for the fidelity and $\Delta$ analysis of 2D topological superconductors and insulators. In our study the parameters of the Hamiltonian and the temperature are treated on equal footing, that is, our parameter space consists of points $(\mu,T)$, where $\mu$ denotes the parameters of the Hamiltonian that drive the quantum phase transition and $T$ denotes the temperature. We calculated the fidelity $F(\rho,\rho')$ and $\Delta(\rho,\rho')$ between the states $\rho(\mu,T)$ and $\rho'=\rho(\mu',T')$ with $\mu'=\mu+\delta\mu$ and $T'=T+\delta T$. We consider $\delta\mu$ and $\delta T$ to be small, such that the points $(\mu,T)$ and $(\mu',T')$ of the parameter space, where we evaluate the fidelity and $\Delta$, are close to each other. In particular, we treat the following two cases: 
\begin{itemize}
\item $\delta\mu=0.01$ and $\delta T=0$, i.e., probing $F$ and $\Delta$ only with respect to the parameters of the Hamiltonian, while keeping the temperature unchanged.
\item $\delta\mu=0$ and  $\delta T=0.01$, i.e., probing $F$ and $\Delta$ only with respect to the temperature, while keeping the parameters of the Hamiltonian unchanged.
\end{itemize}

The expressions for the aforementioned quantities are calculated with respect to the many-body thermal states
\begin{equation} 
\begin{array}{lcl}
\rho & = & e^{-\beta\mathcal{H}}/Z=\Big(\prod_{\mathbf{k}}\exp\big(-\beta\psi_{\mathbf{k}}^{\dagger}H(\mathbf{k})\psi_{\mathbf{k}}\big)\Big)/Z, \\ [2mm]
  Z & = & \tr\{\exp(-\beta \mathcal{H})\}.
\end{array}
\end{equation}
where $\mathbf{k}=(k_{x},k_{y})$, $\psi_{\bf{k}}=(\psi_{1,\bf{k}},\psi_{2,\bf{k}})$ is an array of fermion field (annihilation and/or creation) operators and $H(\mathbf{k})$ is the ``single-particle'' Hamiltonian of the 2D topological insulator/superconductor in momentum space, whose exact form is presented below in the respective sections (note that it is only single-particle if the theory preserves the total charge). In the case of superconductors $\psi_{1,\bf{k}}\equiv \psi_{\uparrow,\bf{k}}$ and $\psi_{2,\bf{k}}\equiv \psi_{\downarrow,-\bf{k}}^{\dagger}$, i.e., $\psi_{\bf{k}}$ is the Nambu spinor and $H(\bf{k})$ is the Bogoliubov-de Gennes Hamiltonian. The occupation number, for each $\bf{k}$, of each of the bands is thermal,
\begin{align*}
n(\varepsilon_{\alpha,\bf{k}})=\frac{1}{e^{\beta \varepsilon_{\alpha,\bf{k}}}+1}, \text{ where } \alpha, \text{ labels the bands, and } \bf{k} \text{ labels the quasi-momenta},
\end{align*}
according to the Fermi-Dirac distribution, as can be derived from a Bogoliubov transformation diagonalizing $\mathcal{H}$. The expression for $\mathcal{H}$, in general, includes the chemical potential $\mu$, and therefore it is included in the expression for $\varepsilon_{\alpha,\bf{k}}$. Generically, it is different from $1$. Only if $\varepsilon_{\alpha,\bf{k}}<0$ (below the Fermi level) it will reach $1$ in the limit $T\to 0$ (i.e., $\beta\to +\infty$).

 For the analytic derivation of the expressions of the fidelity and the quantity $\Delta$, see Appendix~\ref{sec:Analytic-Expression-for-fidelity}.

\subsection{2D topological superconductor}

We consider the lattice Hamiltonian of a chiral p-wave superconductor~\cite{ber:hug:13,vol:99,rea:gre:00,iva:01}
\begin{equation}
\mathcal{H} =\sum_{ij}[-t(c_{i+1,j}^{\dagger}c_{i,j}+c_{i,j+1}^{\dagger}c_{i,j})-\dfrac{1}{2}(\mu-4t)c_{i,j}^{\dagger}c_{i,j}+ S(c_{i+1,j}^{\dagger}c_{i,j}^{\dagger}+i c_{i,j+1}^{\dagger}c_{i,j}^{\dagger})+\text{H. c.}],
\end{equation}
where $c_{i,j}$ and $c_{i,j}^{\dagger}$ are fermionic annihilation and creation operators, respectively. $S$ is the superconducting pairing, $\mu$ is the chemical potential and $t$ is the nearest-neighbor hopping. To simplify, we fix $|S|=t=\dfrac{1}{2}$. We can transform this Hamiltonian in momentum space -- since in real space it is translationally invariant -- and then apply the Bogoliubov-Valatin transformation. We end up with the single-particle Hamiltonian expressed in the Nambu spinor basis 

\begin{equation}
H(\textbf{k})=-\{\sin(k_{y})\sigma_{x}+\sin(k_{x})\sigma_{y}+[\mu-2+\cos(k_{x})+\cos(k_{y})]\sigma_{z}\},	
\end{equation}
with $\sigma_{x,y,z}$ being the Pauli matrices. At $T=0$, the Chern number ($\text{Ch}$) characterizes two distinct non-trivial topological phases~\cite{viy:riv:del:2d:14}: one with $\text{Ch}=1$ for $0<\mu<2$ and one with $\text{Ch}=-1$ for $2<\mu<4$. For $\mu<0$ and $\mu>4$, the system is in a topologically trivial phase ($\text{Ch}=0$).

\begin{figure}[h!]

\subfloat[]{\includegraphics[width=2.2in,height=1.7in]{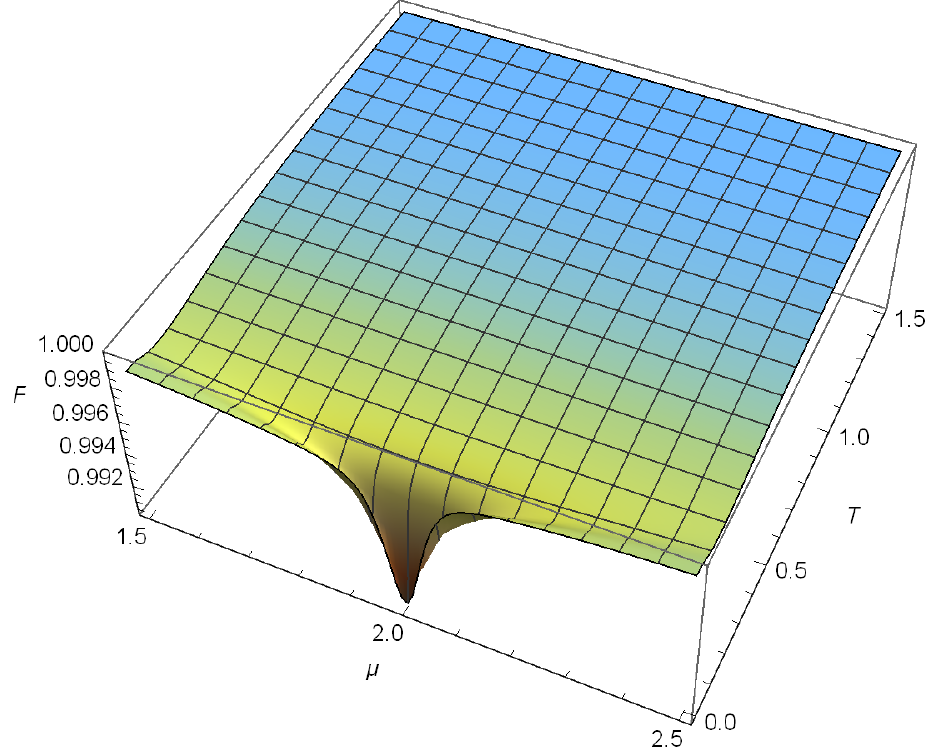}}
\subfloat[]{\includegraphics[width=2.2in,height=1.7in]{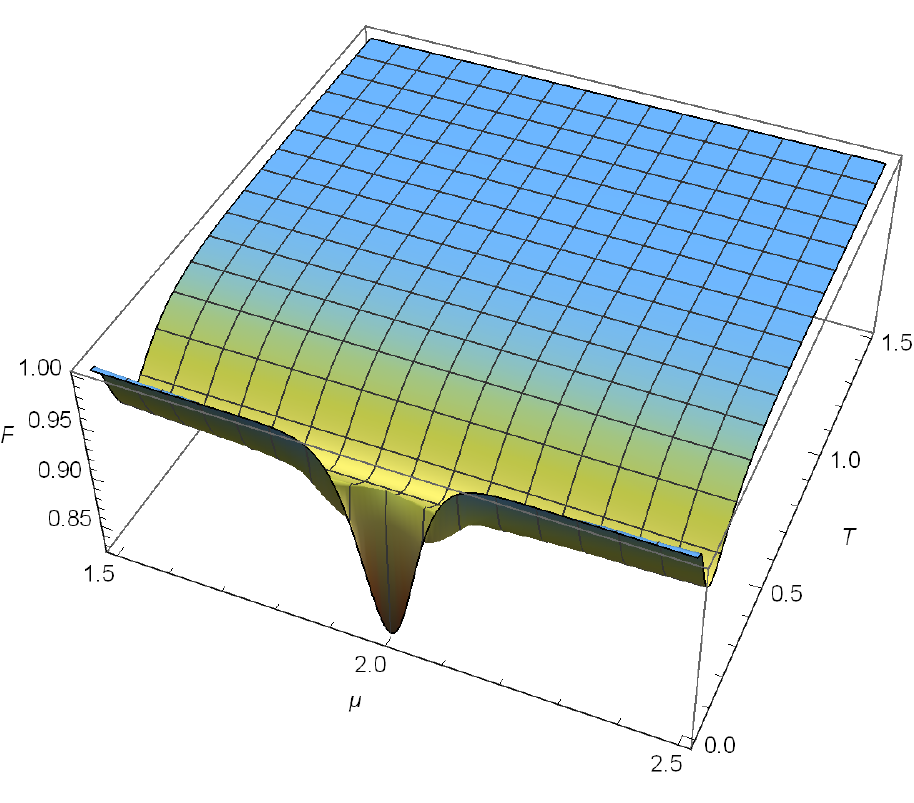}}
\subfloat[]{\includegraphics[width=2.2in,height=1.7in]{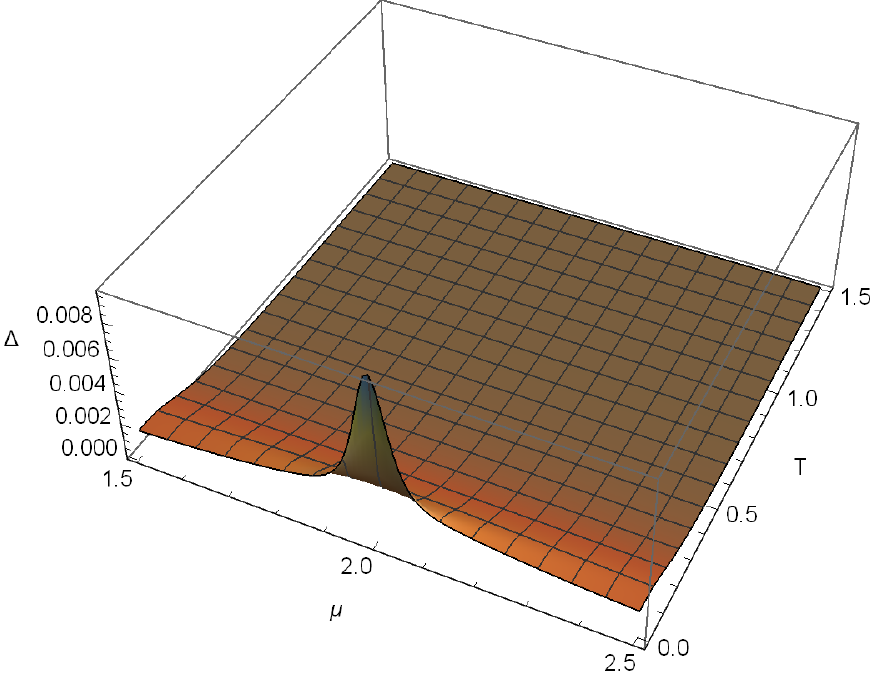}}

\center
\caption{Fidelity for the thermal state $\rho$, (a) when probing the parameter of Hamiltonian $\delta\mu=0.01$, (b) when probing the temperature $\delta T=0.01$, and (c) the quantity $\Delta$ when probing the Hamiltonian parameter $\delta\mu=0.01$, for the 2D topological superconductor.}

\label{fig:Fidelity_TS}
\end{figure}

In FIG.~\ref{fig:Fidelity_TS} 
, we present the plots for the fidelity and $\Delta$, when changing the parameters of the Hamiltonian and the temperature, respectively. We observe that at $T=0$, in both cases the fidelity drops around the gap closing point at $\mu=2$. As we increase the temperature these drops gradually disappear, indicating that the temperature smears out the topological features. Plot (c) shows the behaviour of $\Delta$, for $\delta\mu=0.01$ and $\delta T=0$, i.e., when we change the parameter of the Hamiltonian. At $T=0$ , $\Delta$ becomes non-trivial in the neighborhood of the gap-closing point at $\mu=2$, signalling the topological phase transition, just like the fidelity. For higher $T$, this non-triviality is smeared out, indicating again the absence of finite-temperature phase transitions. For $\delta\mu=0$ and $\delta T=0.01$, i.e, when we only change the temperature, $\Delta=0$ everywhere, as the Hamiltonians commute at different temperatures (we omitted the respective plot). The triviality of $\Delta$, with respect to the change of temperature, can be explained by the fact that it is only sensitive to changes of the eigenbasis, while the fidelity is sensitive to both changes of the spectrum and eigenbasis~\cite{mer:vla:pau:vie:17}.

\subsection{2D topological insulator}

In 2D, a topological insulator can be realized by placing an atom with an internal degree of freedom at each site of a triangular lattice. This model is described by the following Hamiltonian~\cite{sti:pie:fuc:kal:sim:12}
\begin{align}
\mathcal H =\sum_{ij}[c_{i+1,j}^{\dagger}(t_{1}\sigma_{x}+it_{3}\sigma_{z})c_{i,j}+c_{i,j+1}^{\dagger}(t_{1}\sigma_{y}+it_{3}\sigma_{z})c_{i,j}+c_{i+1,j+1}^{\dagger}(t_{2}\sigma_{z})c_{i,j}+\text{H. c.}].
\label{eq:RealSpaceTI_Hamiltonian}
\end{align}

Again, by taking advantage of the translational invariance of the Hamiltonian in real space, we transform it in momentum space, and considering periodic boundary conditions we obtain
\begin{align}
H(\textbf{k}) =2t_{1}\cos(k_{x})\sigma_{x}+2t_{1}\cos(k_{y})\sigma_{y}+\{2t_{2}\cos(k_{x}+k_{y})+2t_{3}[\sin(k_{x})+\sin(k_{y})]\}\sigma_{z}.
\label{eq:H(kx, ky)}
\end{align}

\begin{figure}[h!]

\subfloat[]{\includegraphics[width=2.2in,height=1.7in]{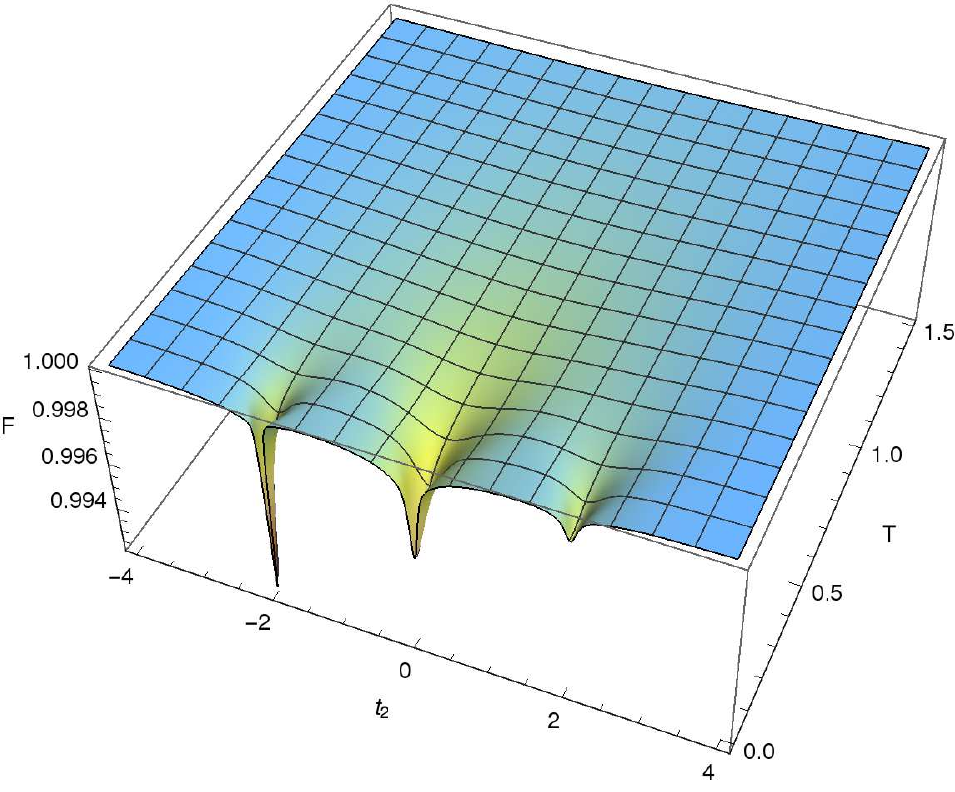}}
\subfloat[]{\includegraphics[width=2.2in,height=1.7in]{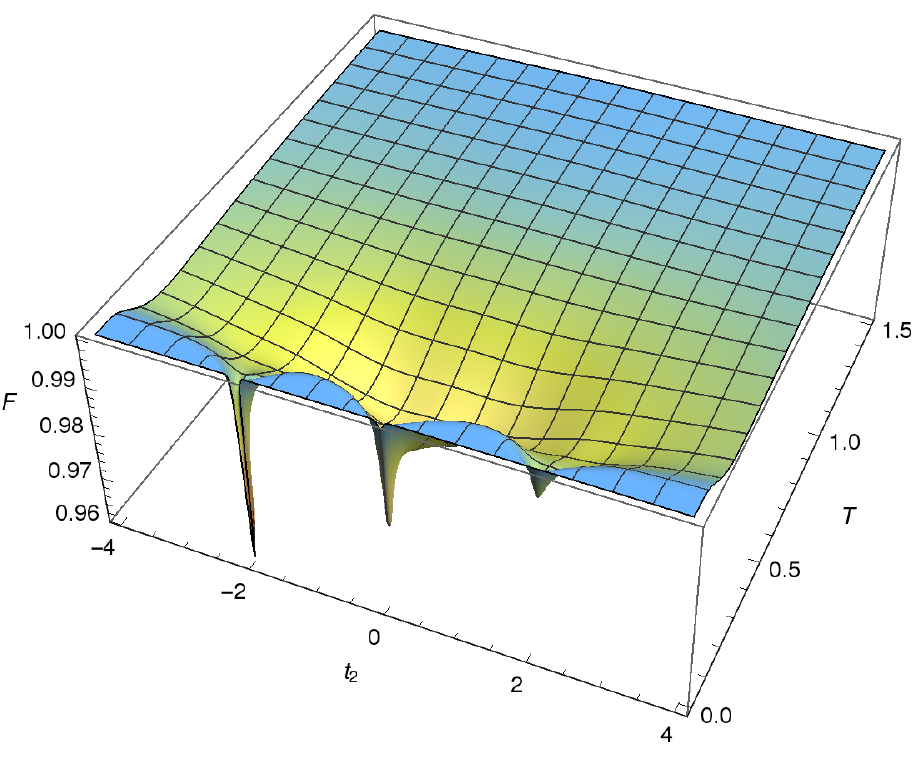}}
\subfloat[]{\includegraphics[width=2.2in,height=1.7in]{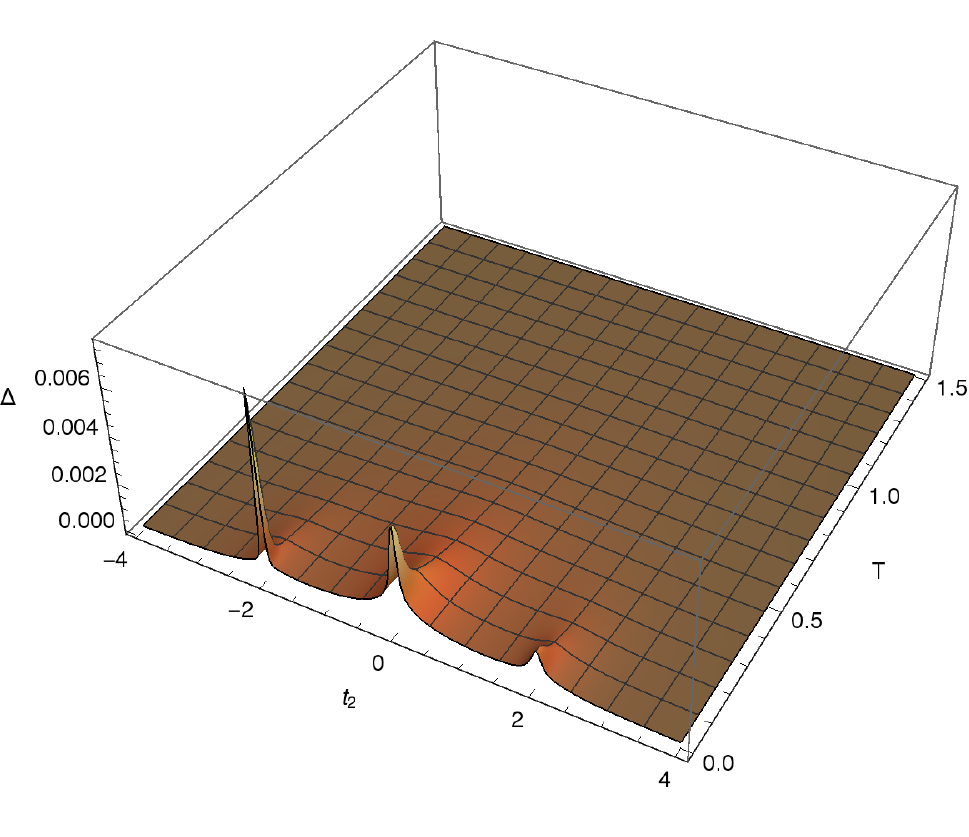}}

\par

\center
\caption{Fidelity for the thermal state $\rho$, (a) when probing the parameter of Hamiltonian $\delta t_{2}=0.01$, (b) when probing the temperature $\delta T=0.01$, and (c) the quantity $\Delta$ when probing the Hamiltonian parameter $\delta t_{2}=0.01$, for the 2D topological insulator.}

\label{fig:Fidelity_TI}
\end{figure}

In order to simplify, we fix $t_{1}=t_{3}=1$, as in~\cite{viy:riv:del:2d:14}.  At $T=0$, by varying the value of the parameter $t_2$, we can find four distinct topologically non-trivial phases with different Chern numbers: $\text{Ch}=\pm2$ for $t_{2}\lessgtr\mp2$ and $\text{Ch}=\pm1$ for $\mp2\lessgtr t_{2}\lessgtr0$. In FIG.~\ref{fig:Fidelity_TI}, we show the respective plots for the fidelity and $\Delta$. Similarly to the previous case of the 2D topological superconductor, at zero temperature we observe drops of the fidelity at the gap-closing points $t_{2}=-2,0,+2$, which signal the topological phase transitions (plots (a) and (b)). When we increase the temperature, the fidelity goes gradually back to 1, indicating the absence of finite-temperature phase transitions.

Finally, in plot (c), we observe that $\Delta$, when we vary the parameter of the Hamiltonian ($\delta t_{2}=0.01, \delta T=0$), becomes non-trivial around the gap-closing points at $T=0$, indicating the topological phase transitions, which are gradually smeared out as temperature increases. In the case where we only change the temperature ($\delta t_{2}=0, \delta T=0.01$) $\Delta$ is, again, zero everywhere, due to the fact the change of temperature only affects the spectrum of the Hamiltonian (we also omitted the respective plot).

Regarding the relation between the two quantities analyzed, note that the quantity $\Delta$ is manifestly trivial for the case of mutually commuting Hamiltonians. Thus, knowing the Hamiltonians do commute, $\Delta =0$ brings no new information about the system's behavior. Since in our case Hamiltonians do not depend on the temperature, the behavior of $\Delta$ when only the temperature is varied tells us nothing about the existence of temperature-driven transitions, which are, at least in principle, possible even in those cases. Nevertheless, the fidelity, which is sensitive to {\em both} the change of Hamiltonian's eigenvectors, as well as its eigenvalues, can signal the existence of thermally-driven phase transitions. Thus, one should consider the results of the fidelity as more general than those of the quantity $\Delta$. Nevertheless, the possible nontrivial behavior of $\Delta$ gives additional and more refined information than the fidelity can provide.

 \section{Absence of finite temperature phase transitions: Fidelity Susceptibility}
 
 In this section, we perform both analytical and numerical analysis of the fidelity susceptibility in the thermodynamic limit, providing an explicit quantitative criterion for the existence of phase transitions -- if it diverges, there exists a phase transition, otherwise there is no phase transition. At the end, we also compute the respective critical exponents.
 
 To fix notation, the Hamiltonian for the class of two-band systems considered can be cast in the following form
 \begin{eqnarray}
 \mathcal{H}=\int_{\text{B.Z.}} \frac{d^2 k}{(2\pi)^2} \psi^{\dagger}_{\bf{k}}H(\bf{k})\psi_{\bf{k}}, \text{ with } H(\bf{k})=d^{\mu}(\bf{k};M)\sigma_{\mu},
 \end{eqnarray}
 where $\psi_{\bf{k}}$ is an array of fermion field operators, $\sigma_{\mu}$ are the usual Pauli matrices, $M$ is some parameter (e.g., a hopping amplitude) which drives the topological phase transition and the Einstein summation convention is assumed. 
 
\begin{figure}[h]
\centering
\includegraphics[scale=0.66]{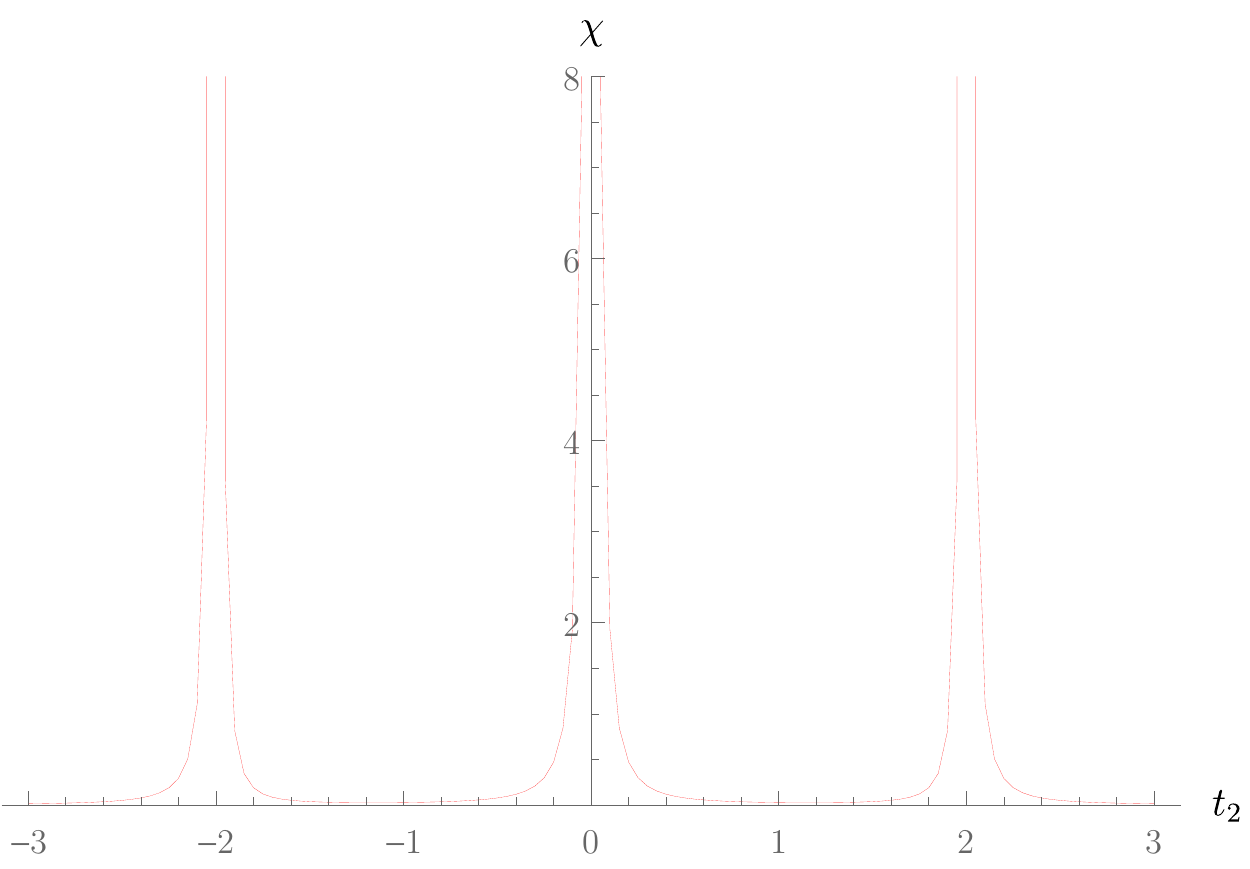}
\caption{The thermodynamic limit of the fidelity susceptibility at zero temperature for the $2D$ topological insulator model. The singularities appear precisely at the critical points of quantum phase transition where the gap closes.}
\label{fig:fidsusc_zero_temp_ti}
\end{figure}

 The absence of finite temperature phase transitions in the systems considered can be justified through a careful analysis of the fidelity susceptibility $\chi_{MM}$. In doing so there are two limits to be considered: the thermodynamic limit, in which one considers the system on a finite lattice with periodic boundary conditions and then takes the number of its points to infinity, and the zero temperature limit. The order in which one takes these two limits is important and, in fact, it produces different results. The fidelity susceptibility is defined through the formula
 \begin{eqnarray}
\frac{1}{2} \chi_{MM}\delta M^2=-\frac{\log \big[F(\rho(M),\rho(M+\delta M))\big]}{N^2 },
  \end{eqnarray}
where the fidelity is taken with respect to infinitesimally close thermal states, of a fixed temperature, $\rho(M)$ and $\rho(M+\delta M)$ with respect to a variation of the parameter $M$, namely $M\to M+\delta M$, and $N^2$ is the number of lattice sites ($N$ sites on each of the two directions in real space). It is the pullback by $\rho:M\mapsto \rho(M)$ of the Bures metric (appropriately normalized to take the thermodynamic limit). 
  Taking the zero temperature limit and then the thermodynamic limit yields the following expression for the fidelity susceptibility
 \begin{eqnarray}
 	\chi_{MM}=\int_{\text{B.Z.}} \frac{d^2 k}{(2\pi)^2} \frac{1}{4}\delta_{\mu\nu}\frac{\partial n^{\mu}}{\partial M}\frac{\partial n^{\nu}}{\partial M},
 \label{eq:zero_temp_fidsusc}
 \end{eqnarray}
 where $n^{\mu}=d^{\mu}/|d|$ is the normalized Bloch vector. The previous formula manifests the pure state character of the zero temperature limit of the thermal state. Indeed, it is the integral over the first Brillouin zone of the pullback of the Fubini-Study metric on the Bloch sphere by family of maps $M\mapsto n^{\mu}(M)=d^{\mu}(M)/|d(M)|$ parametrized by the quasimomentum $\bf{k}\in \text{B.Z}$.  The above integral can be numerically evaluated, and the results for the case of topological insulator, given by the Hamiltonian~\eqref{eq:RealSpaceTI_Hamiltonian}, are presented in Fig.~\ref{fig:fidsusc_zero_temp_ti}. The fidelity susceptibility diverges precisely at the gapless points that define quantum criticality, and are given by $M\equiv t_2=0,\pm 2$.
 
 By taking finite temperature expression for the fidelity susceptibility, and performing the thermodynamic limit, we obtain
 \begin{eqnarray}
 \chi_{MM}=\int_{\text{B.Z.}} \frac{d^2 k}{(2\pi)^2}\Big[\frac{1}{4}\Big(\frac{\cosh(\beta E(\bf{k}))-1}{\cosh(\beta E(\bf{k}))}\delta_{\mu\nu}\frac{\partial n^{\mu}}{\partial M}\frac{\partial n^{\nu}}{\partial M}+\frac{1}{\cosh\big(\beta E(\bf{k})\big)+1}\Big(\frac{\partial E (\bf{k})}{\partial M}\Big)^2\Big)\Big],
 \label{eq:finite_temp_fidsusc}
\end{eqnarray}
where $E(\bf{k})=|d(\bf{k})|$ and $\beta=1/T$ is the inverse of the temperature (see Appendix~\ref{Appendix C} for the derivation). One is then interested in taking the limit $T\to 0$ of the previous expression. A naive approach would give the same as Eq.~\eqref{eq:zero_temp_fidsusc}. However, if one takes into consideration the existence of the gapless points, the result is different. While the expression of Eq.~\eqref{eq:zero_temp_fidsusc} yields singularities at the critical points, the later Eq.~\eqref{eq:finite_temp_fidsusc} is perfectly regular. The reason is the second term coming from the variation of the energy which seems to magically cure the divergence.
When starting with the Boltzmann-Gibbs states, taking the thermodynamic limit and then taking the zero temperature limit of the fidelity susceptibility, the second term appears and is relevant specially in the neighbourhood of gap closing points. This term is precisely the difference between taking the zero temperature limit and then the thermodynamic limit and vice-versa. For generic non-zero values of $E(\bf{k})$, the limit $\beta\to +\infty $ of the integrand function yields,
\begin{eqnarray}
\frac{1}{4}\delta_{\mu\nu}\frac{\partial n^\mu}{\partial M}\frac{\partial n^\nu}{\partial M},
\end{eqnarray}
as expected.
However, near the gap closing points this is not true. In this case, we take $\beta$ finite and start by expanding in $E(\bf{k})$,
\begin{align*}
&\frac{1}{8}(\beta E(\bf{k}))^2\delta_{\mu\nu}\frac{\partial n^\mu}{\partial M}\frac{\partial n^\nu}{\partial M}+\frac{1}{8}\beta^2 \Big(\frac{\partial E(\bf{k})}{\partial M}\Big)^2 +\mbox{O}\big((\beta E(\bf{k}))^4\big)\\
&=\frac{1}{8}\beta^2\delta_{\mu\nu}\frac{\partial(E n^{\mu})}{\partial M}\frac{\partial(E n^{\nu})}{\partial M}+\mbox{O}\big((\beta E(\bf{k}))^4\big)\\
&=\frac{1}{8}\beta^2\delta_{\mu\nu}\frac{\partial d^{\mu}}{\partial M}\frac{\partial d^{\nu}}{\partial M}+\mbox{O}\big((\beta E(\bf{k}))^4\big).
\end{align*}
The last expression is way too raw to take the zero temperature limit. Notice that we lost the denominators in this approximation. The denominators in both terms are roughly the same both in the $\beta\to+\infty$ while keeping $E(\bf{k})$ finite but non zero and $E(\bf{k})\to 0$ while keeping $\beta$ finite. We have the inequality,
\begin{eqnarray}
\frac{1}{2\cosh(x)}< \frac{1}{\cosh(x)+1}< \frac{1}{\cosh(x)}, \text{ for all } x\in \mathbb{R}.
\end{eqnarray}
Therefore, we can safely upper bound the integrand by
\begin{eqnarray}
\frac{1}{4}\frac{\beta^2}{\cosh\big(\beta E(\bf{k})\big)}\delta_{\mu\nu}\frac{\partial d^{\mu}}{\partial M}\frac{\partial d^{\nu}}{\partial M}+\mbox{O}\big((\beta E(\bf{k}))^4\big)
\end{eqnarray}
in a neighbourhood of the gap closing points. Notice that the integrand will be positive or zero since it is a pullback of a Riemannian metric. However, if we now take the zero temperature limit, this upper bound vanishes. Since the singularities at zero temperature were coming from the gap closing points this shows why there are no finite temperature phase transitions in the models discussed.

\begin{figure}[h!]

\subfloat[T = 0.009]{\includegraphics[width=1.8in,height=1.3in]{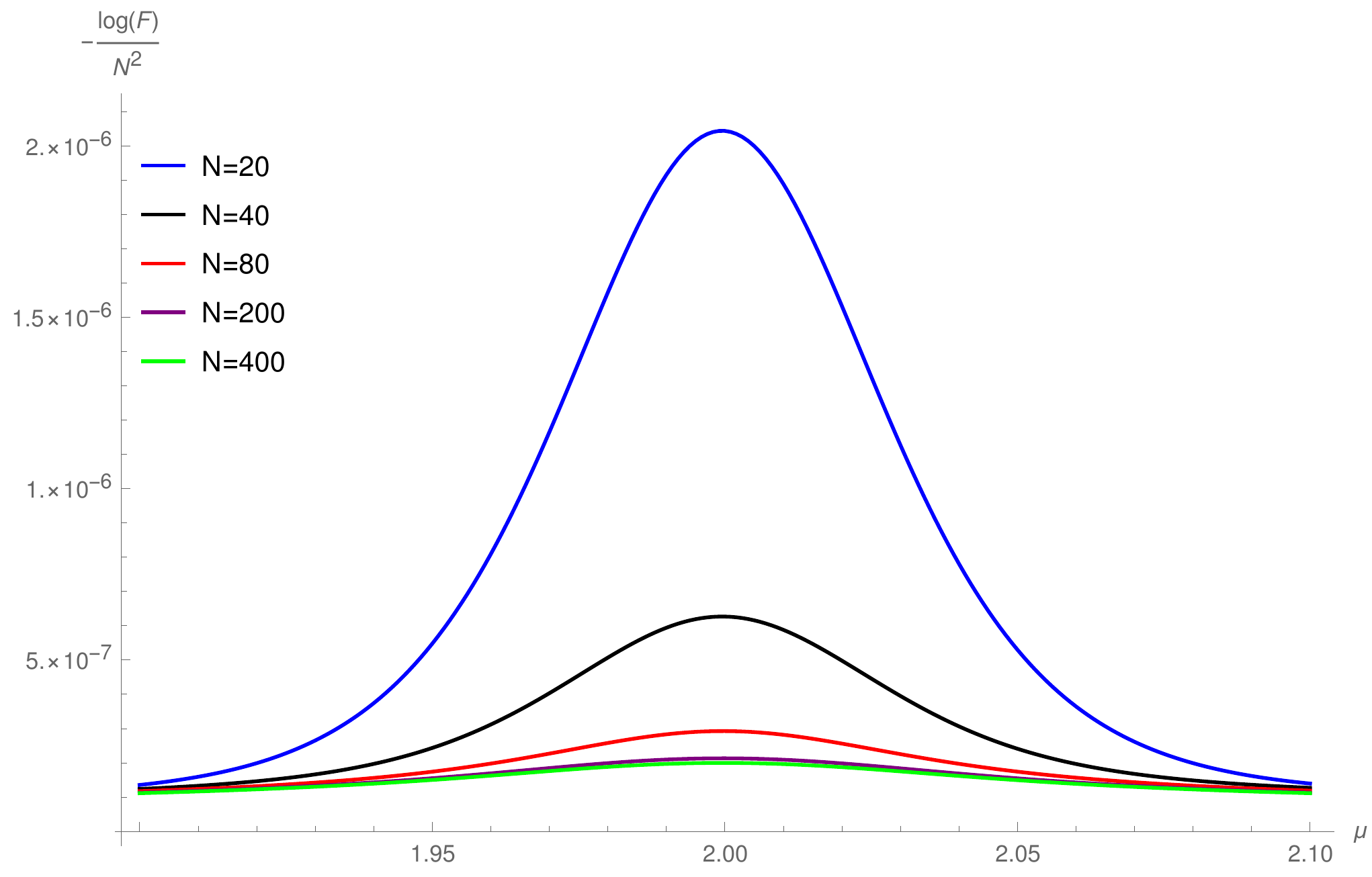}}
\subfloat[T = 0.003]{\includegraphics[width=1.8in,height=1.3in]{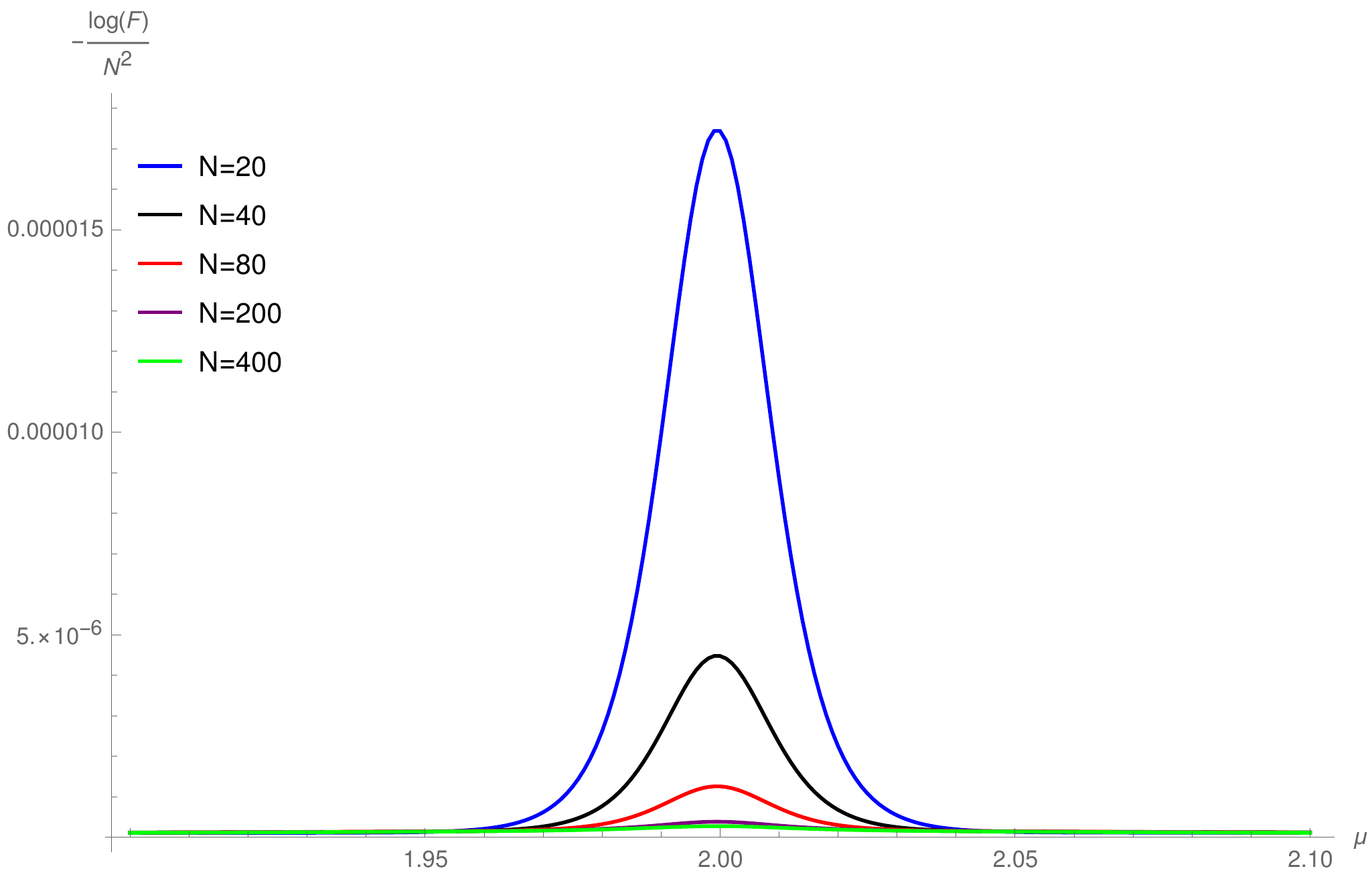}}
\subfloat[T = 0]{\includegraphics[width=1.8in,height=1.3in]{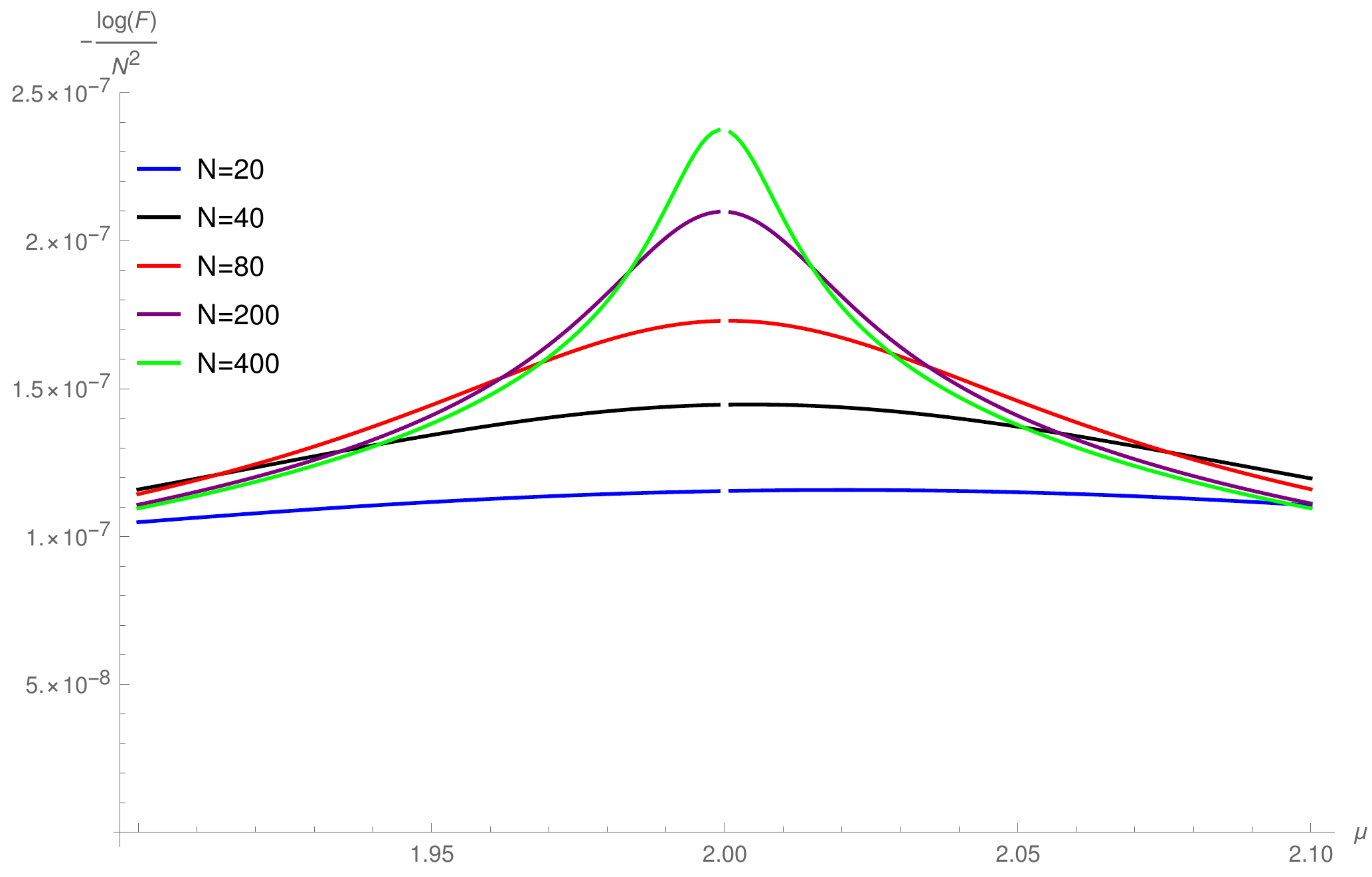}}

\par

\subfloat[T = 0.009 ]{\includegraphics[width=1.8in,height=1.3in]{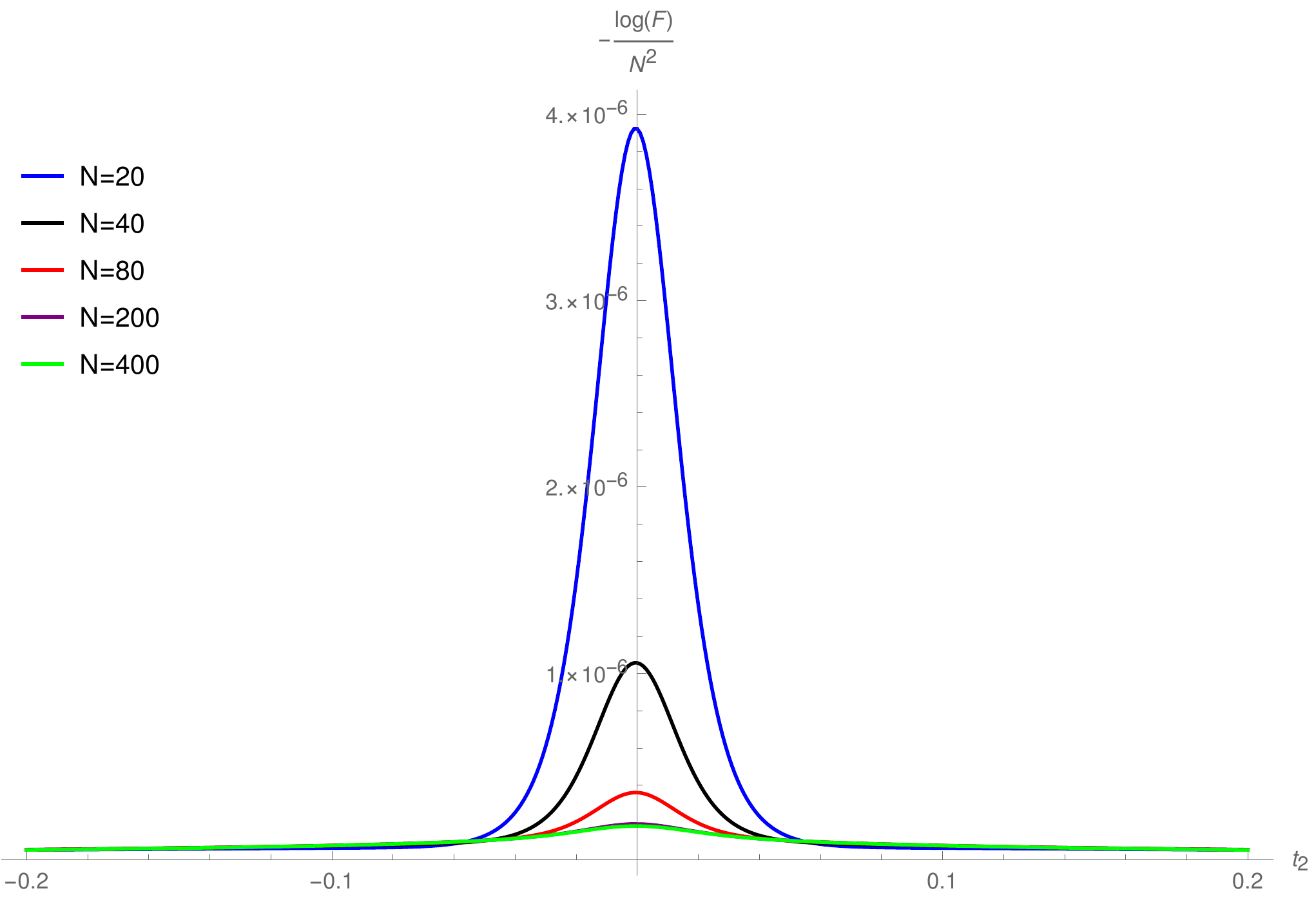}}
\subfloat[T = 0.003]{\includegraphics[width=1.8in,height=1.3in]{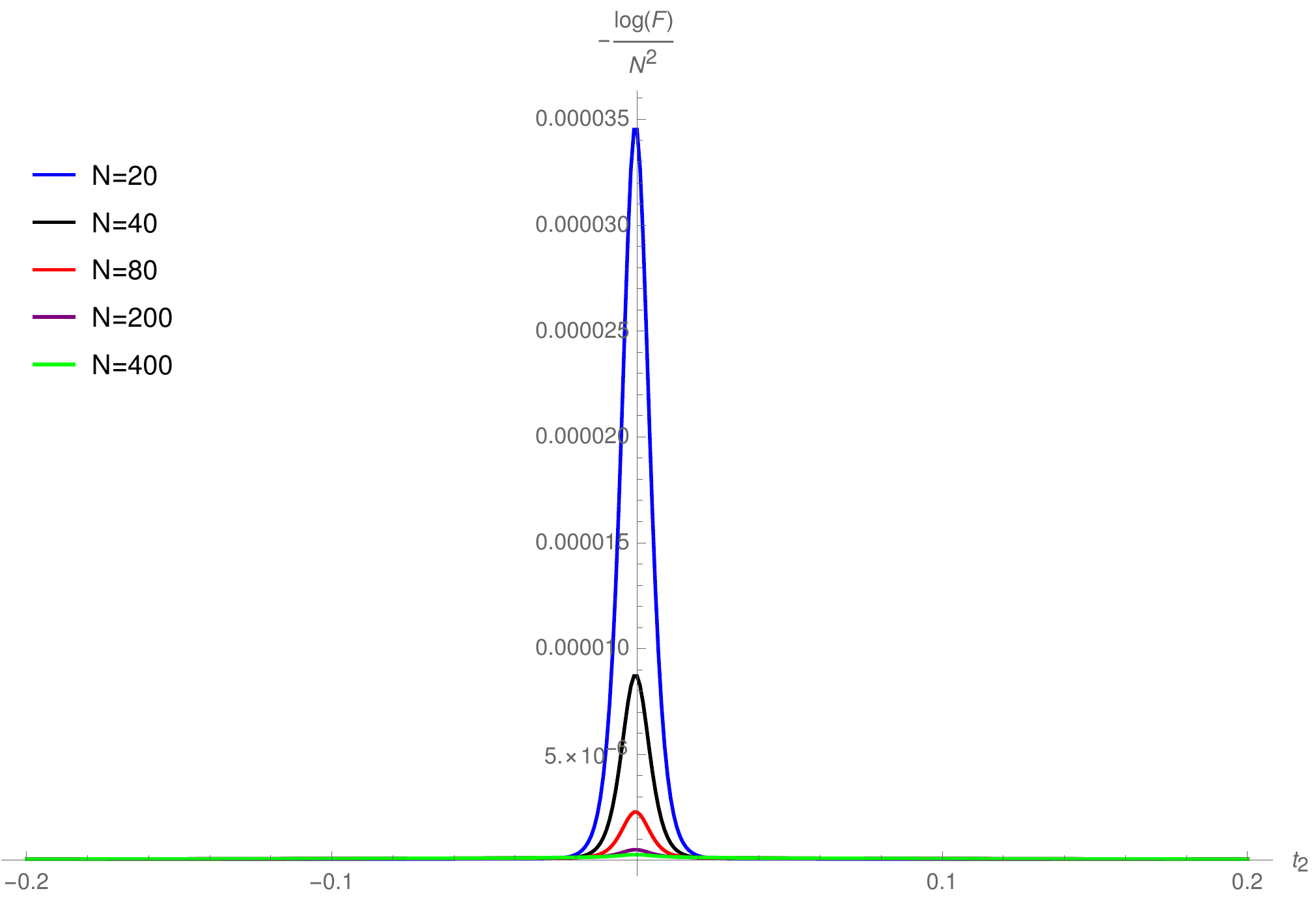}}
\subfloat[T = 0]{\includegraphics[width=1.8in,height=1.3in]{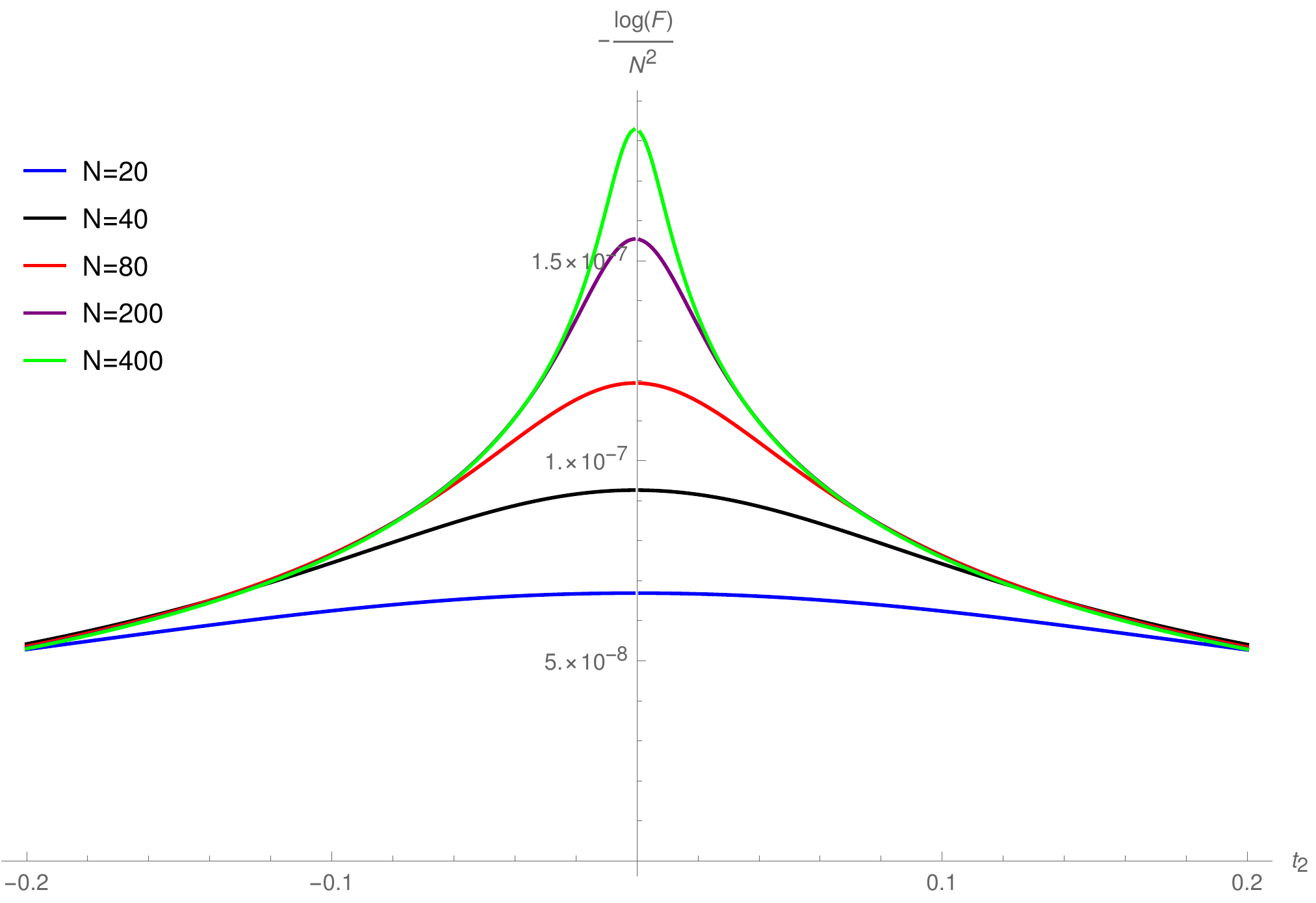}}
\par

\caption{$-\log F /N^2$ for different system size against parameter that control quantum phase transition. (a), (b) and (c) are for 2D topological superconductor, while (d), (e) and (f) are for 2D topological insulator.}
\label{fig: LogFidelity}
\end{figure}

The above analytical analysis is confirmed by numerical evaluation of $-\log F /N^2$ for both the topological superconductor and insulator, presented at Fig.~\ref{fig: LogFidelity}.

The absence of finite-temperature phase transitions means that at temperature different from zero, the fidelity susceptibility $\chi$ does not signal phase transitions, as it does for temperature zero. By changing the parameters of the Hamiltonian at $T=0$, we are able to close the energy gap and have a phase transition from a topological to a trivial phase and/or vice versa, which is detected by the fidelity susceptibility due to the enhanced distinguishability of the states close to the point of phase transition. For finite temperatures ($T>0$), however, we do not observe any changes of $\chi$, which implies the non-existence of phase transitions, driven neither by the parameters of the Hamiltonian nor by the temperature. In this context, one cannot talk about \textit{``a crossover from the zero-temperature topological phase to the infinite-temperature non-topological phase''}, since on one hand, already at $T=0$, we have topological and trivial phases, and on the other hand, the fidelity susceptibility detects the phase transitions between them, which do not exist for $T>0$.

\begin{figure}[h!]
\subfloat[The fidelity susceptibility $\chi_{MM}$ as a function of $M$ in a neighbourhood of $M\equiv \mu=0$.]{\includegraphics[scale=0.45]{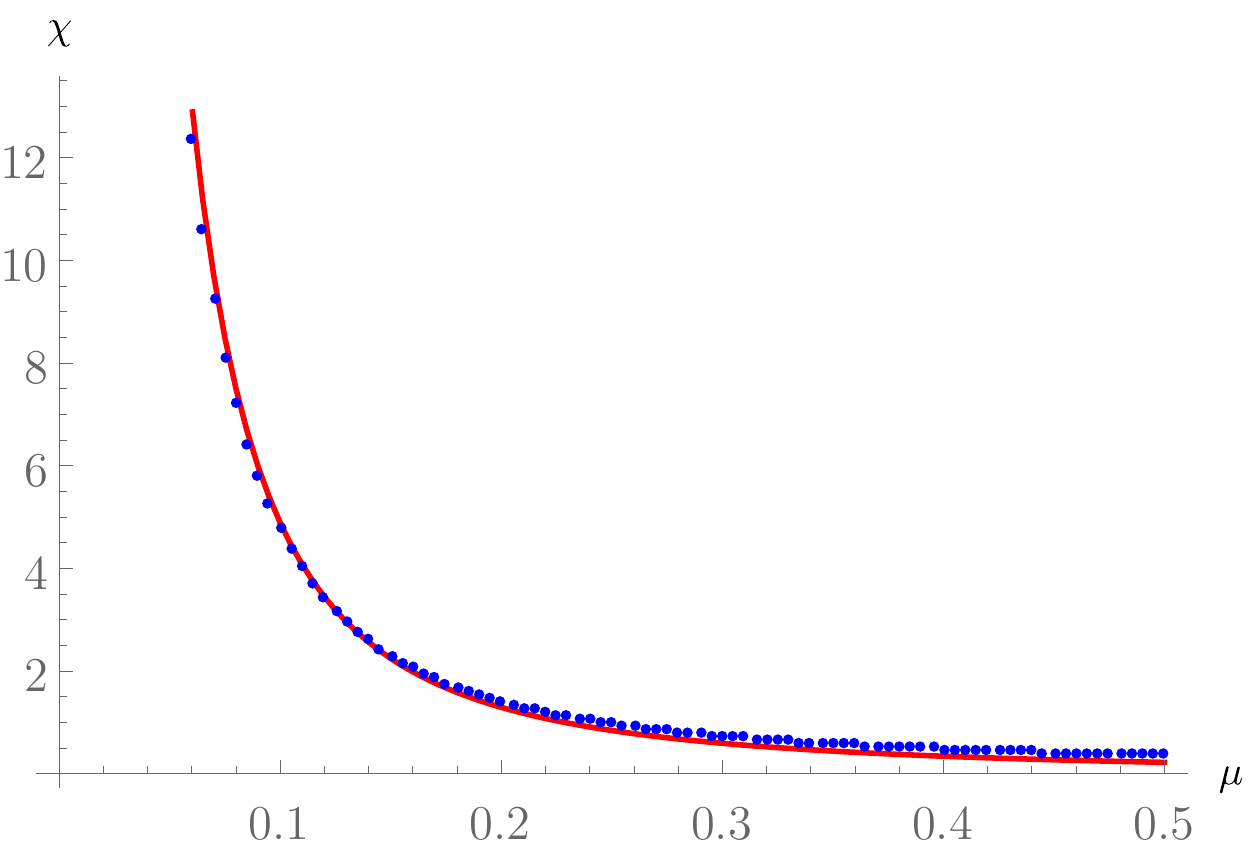}}
\qquad
\subfloat[The fidelity susceptibility $\chi_{MM}$ as a function of $(M-2)$ in a neighbourhood of $M\equiv \mu=2$.]{\includegraphics[scale=0.45]{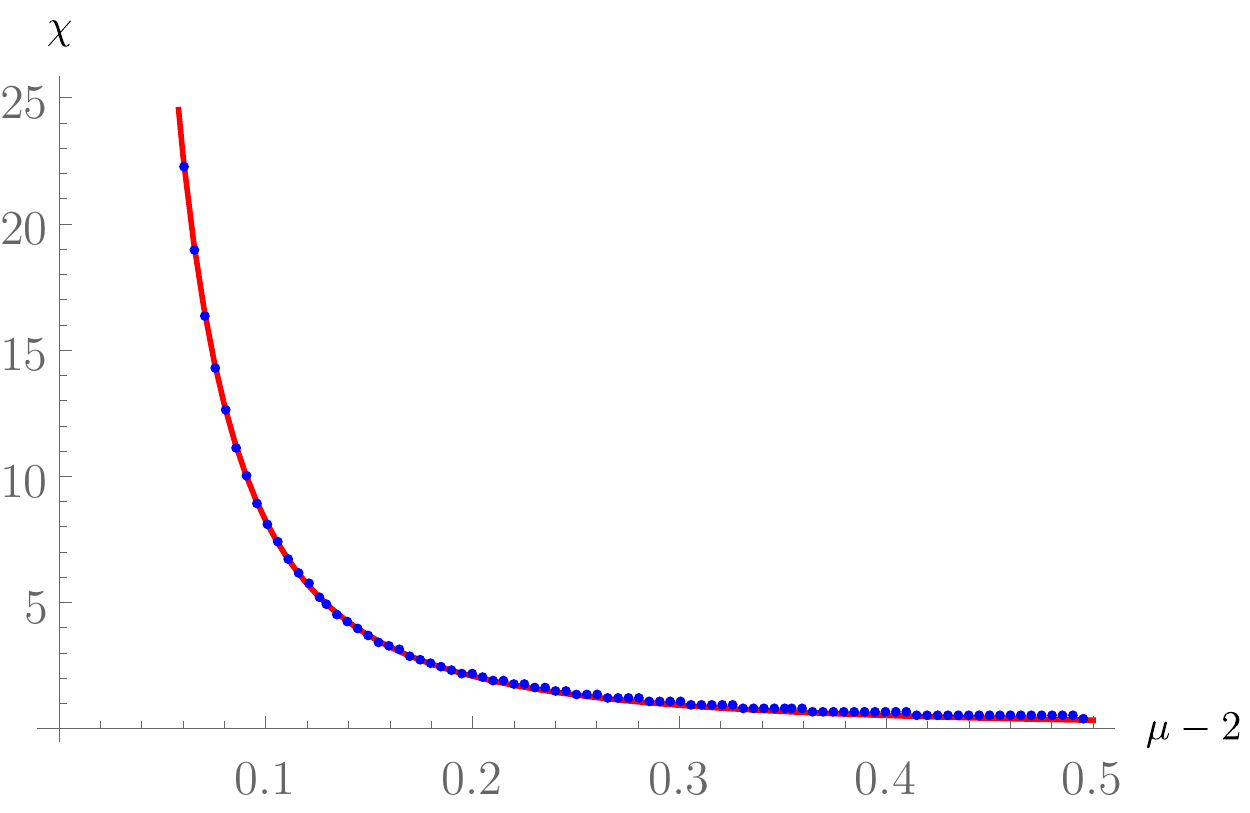}}
\centering
\caption{The fidelity susceptibility $\chi_{MM}$ for the topological superconductor, as a function of $(M-M_c)$. The points are the result of numerical integration. The red curve is the result of the fit.}
\label{fig:crit_exp_tsc}
\end{figure}

\begin{figure}[h!]
\subfloat[The fidelity susceptibility $\chi_{MM}$ as a function of $M$ in a neighbourhood of $M\equiv t_2=0$.]{\includegraphics[scale=0.45]{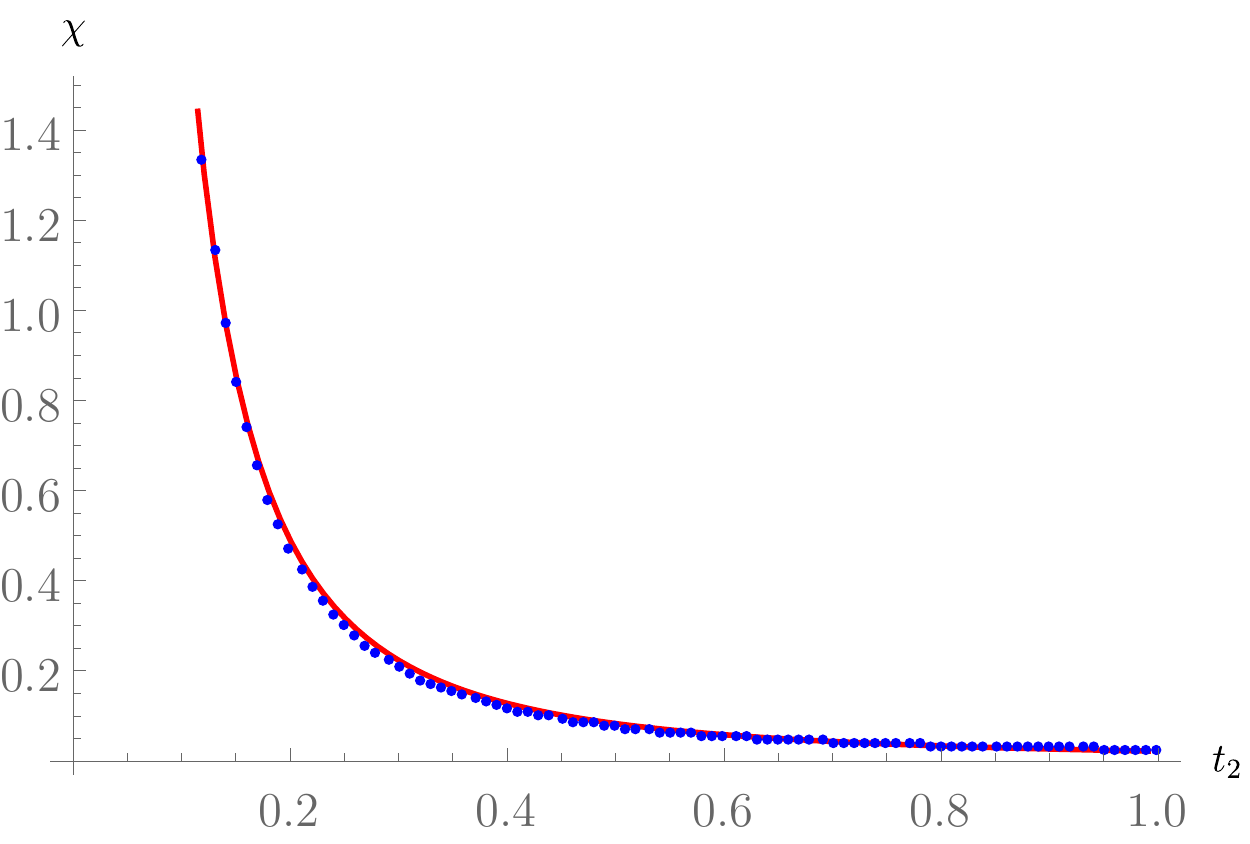}}
\qquad
\subfloat[The fidelity susceptibility $\chi_{MM}$ as a function of $(M-2)$ in a neighbourhood of $M\equiv t_2=2$.]{\includegraphics[scale=0.45]{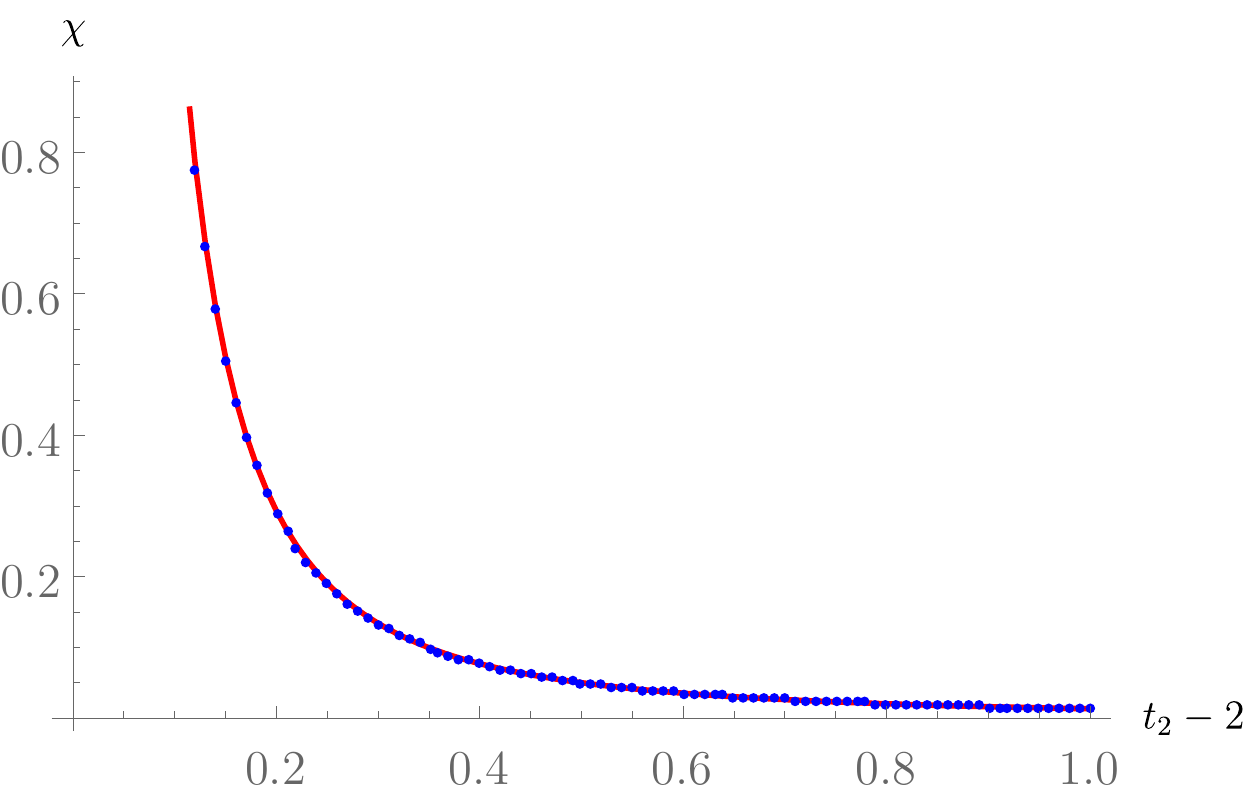}}

\centering
\caption{The fidelity susceptibility $\chi_{MM}$ for the topological insulator, as a function of $(M-M_c)$. The points are the result of numerical integration. The red curve is the result of the fit.}
\label{fig:crit_exp_ti}
\end{figure}

Finally, one can perform the asymptotic analysis of the fidelity susceptibility in the vicinity of the critical points to obtain the critical exponents. For the topological superconductor, we performed a least squares fit to the susceptibility to a power law $a M^{-b}$ in a neighbourhood of $M\equiv \mu=0$ and found $a=0.05906$ and $b=-1.91816$. Furthermore, we also performed a least squares fit to the susceptibility to a power law $a (M-2)^{-b}$ in a neighbourhood of $M\equiv \mu=2$ and found $a=0.08554$ and $b=1.98329$, see FIG.~\ref{fig:crit_exp_tsc} for both cases. Analogously, for the topological insulator, we performed a least squares fit to the susceptibility to a power law $a M^{-b}$ in a neighbourhood of $M\equiv t_2=0$ and found $a=0.02126$ and $b=1.94055$. Additionally, we also performed a least squares fit to the susceptibility to a power law $a (M-2)^{-b}$ in a neighbourhood of $M\equiv t_2=2$ and found $a=0.01309$ and $b=1.93162$, see FIG.~\ref{fig:crit_exp_ti} for both cases. The critical exponents of the fidelity susceptibility of these quantum phase transitions all appear to be around the value $b=1.9$, both for the topological insulator model and for the superconductor.

\section{Spectrum and Edge states for the Topological Insulator}

\begin{figure}[h]
\center
\includegraphics[scale=0.3]{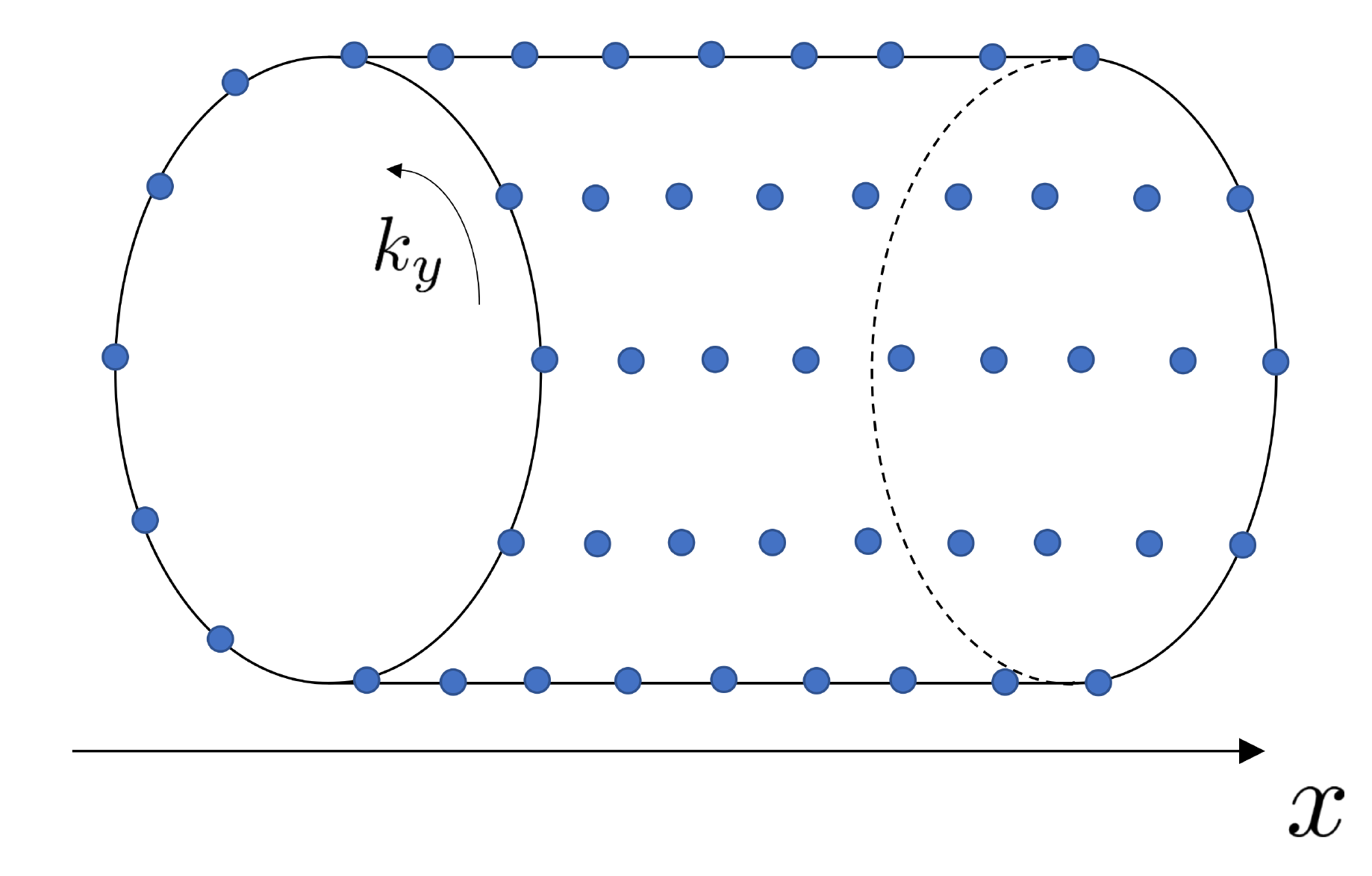}
\caption{To study the edge states, we consider the system on a cylinder topology. The quasi-momentum winds around the periodic direction and the dots denote lattice sites.}
\label{Fig:cylinder}
\end{figure}

Consider the Hamiltonian given by Eq.\eqref{eq:RealSpaceTI_Hamiltonian}, defined on a 2D lattice, describing a topological insulator. We consider the $x$-direction to be finite, with $N$ sites, and with open boundary conditions. For the $y$-direction we take periodic boundary conditions, thus allowing for a partial Fourier transformation along this direction. Topologically, the system is a cylinder, see FIG.~\ref{Fig:cylinder}.

\begin{figure}[h!]
\centering
\includegraphics[scale=0.33]{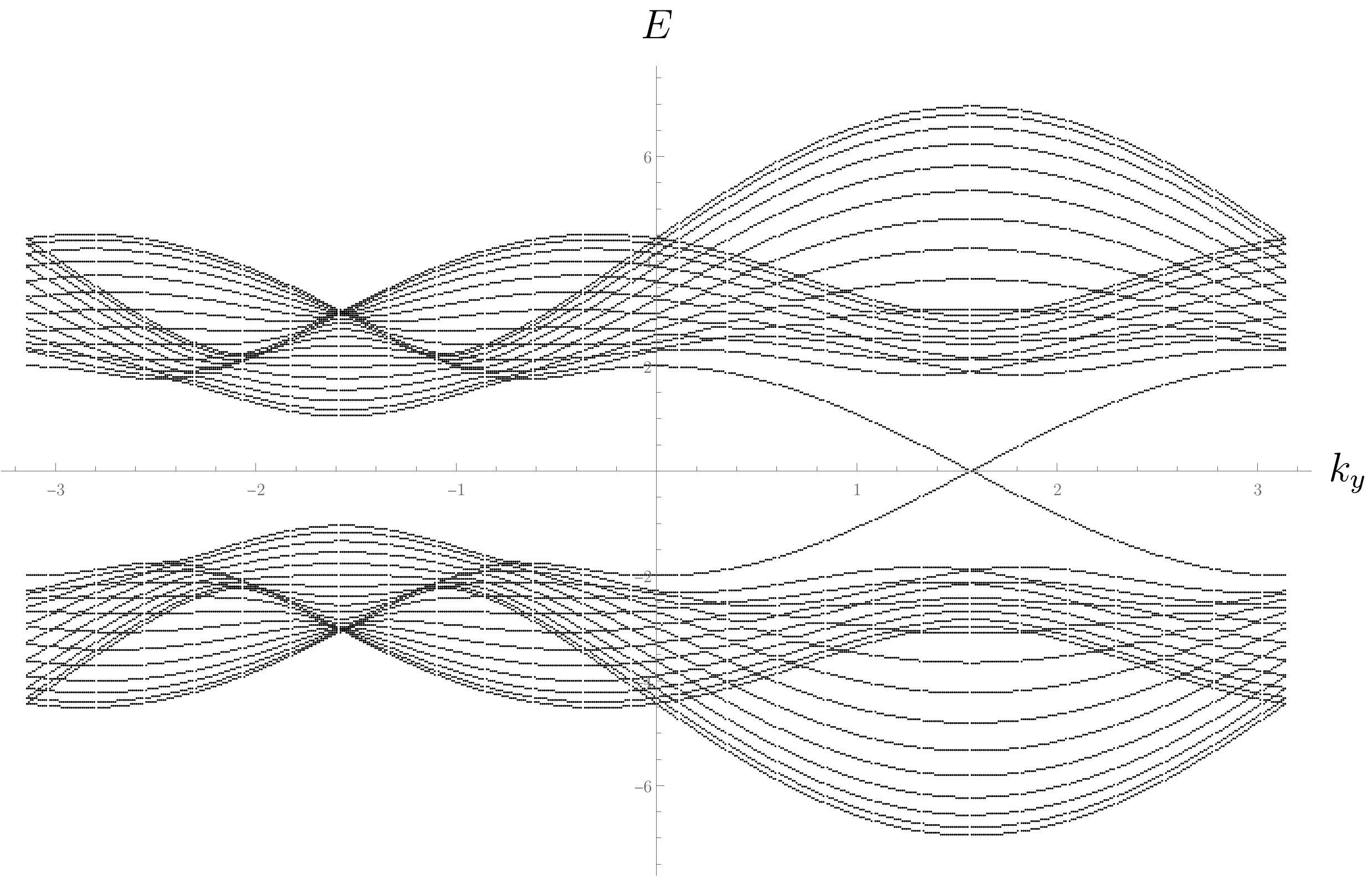}
\caption{Spectrum of the $2$D TI, for $t_2=1.5$ in the cylinder topology. The size of the lattice in the $x$ direction is $20$ sites.}
\label{fig:TI spectrum non-trivial phase cylinder topology}
\end{figure}

The 2D lattice Hamiltonian then reduces to a family of 1D Hamiltonians parametrized by the quasi-momentum $k_{y}$. More concretely, we take the Hamiltonian in Eq.~\eqref{eq:RealSpaceTI_Hamiltonian} and we perform the Fourier transform  in the $j$ index, i.e., along the $y$-axis. The fermion creation and annihilation operators in real space, described by the lattice coordinates $(i,j)$, are then expanded in partial Fourier series 
\begin{align*}
c_{i,j}^{^{\dagger}} & =\frac{1}{\sqrt{N_{y}}}\sum_{k_{y}}\exp(+ik_{y}j)c_{i}^{\dagger}(k_{y}),\\
c_{i,j} & =\frac{1}{\sqrt{N_{y}}}\sum_{k_{y}}\exp(-ik_{y}j)c_{i}(k_{y}),
\end{align*}
where $k_y$ is the quasi-momentum along the $y$-direction. The Hamiltonian can then be written in terms of the partial Fourier modes (see Appendix~\ref{sec:Hamiltonian-transformations}),
\begin{align}
\mathcal{H}= \sum_{k_{y}} \mathcal H (k_{y}) =  \sum_{k_y}\sum_{i,j}c_{i}^{\dagger}(k_{y})H_{i,j}(k_{y})c_{j}(k_{y}),
\end{align}
with $H_{i,j}(k_{y})$ as given in  Eq.~\eqref{eq:H_ij(k_{y})}. For each $k_{y}$, the $N\times N$ matrix $H(k_y)=[H_{i,j}(k_{y})]$ describes a 1D system. Now the energy eigenstates are classified into bulk states and edge states. The bulk states are delocalized, while the edge states are exponentially localized along the $x$-direction, either on the left ($i=1$) or on the right ($i=N$). If the bulk of the system is in a topologically non-trivial phase and we find energy eigenstates within the gap, by the bulk-to-boundary correspondence principle, they should be edge states cancelling the anomaly coming from the bulk. Therefore, when plotting $E(k_y)$ for $t_2$ in a topologically non-trivial phase, we find zero-energy edge states (see FIG.~\ref{fig:TI spectrum non-trivial phase cylinder topology} for $N=20$). The effects of temperature on these states can be studied by considering the thermal states $\rho=\exp\big(-\beta(\mathcal{H}-\mu n)\big)/Z$, where $\beta$ is the inverse of the temperature, $n$ is the total number operator and $\mu$ is the chemical potential. In the topologically trivial case the spectrum of the system decomposes into two sets of bands separated by a gap. In the zero temperature limit and zero chemical potential, the lower bands, i.e. the bands below the Fermi level, are occupied and $\rho$ becomes a projector onto this state.

\begin{figure}[h!]
\subfloat[Occupation number in the bulk, at $T=1/10$.]{\includegraphics[scale=0.20]{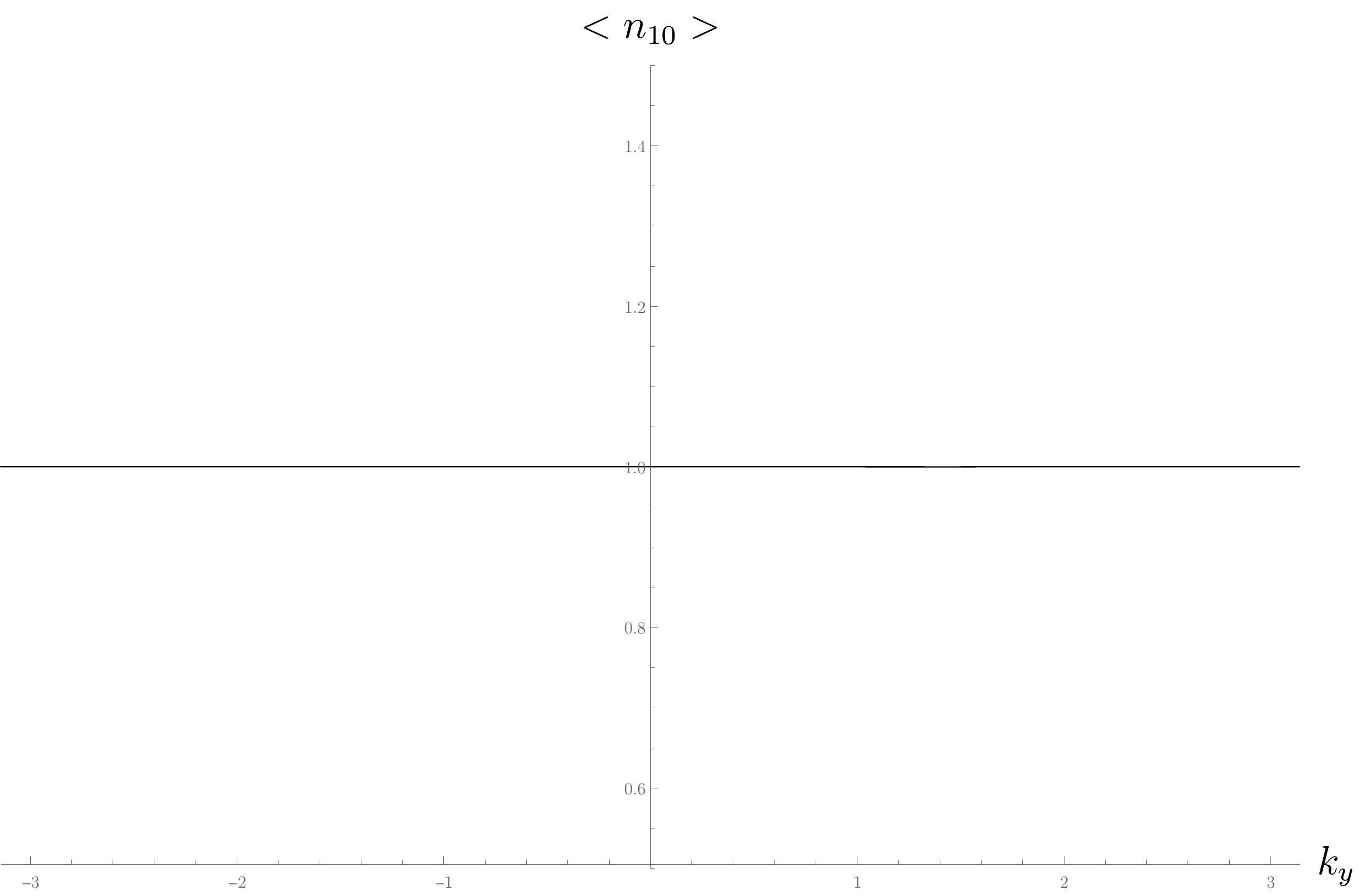}}
\qquad
\subfloat[Occupation number at the left edge, at $T=1/10$.]{\includegraphics[scale=0.20]{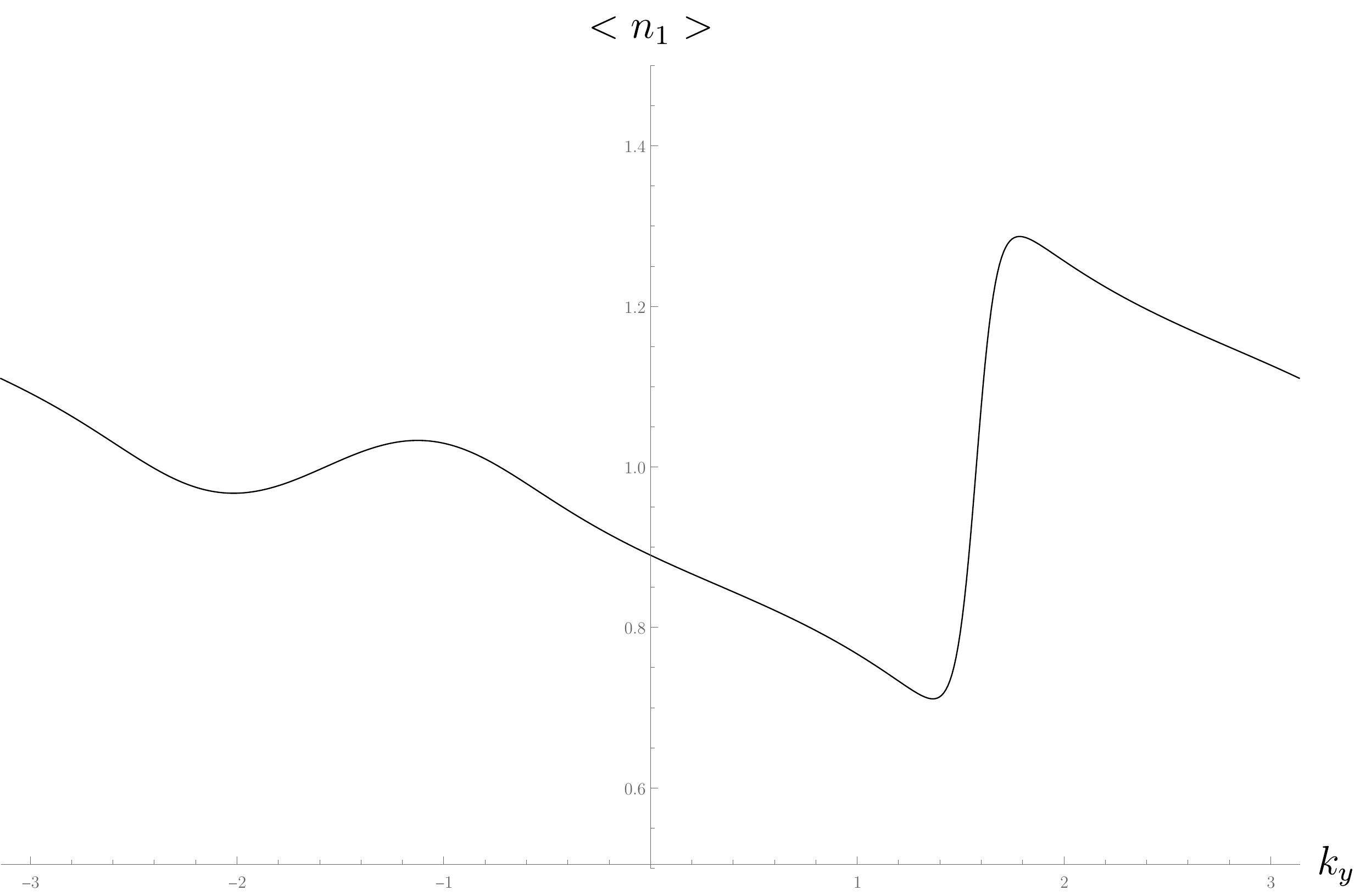}
}
\qquad
\subfloat[Occupation number at the right edge, at $T=1/10$.]{\includegraphics[scale=0.20]{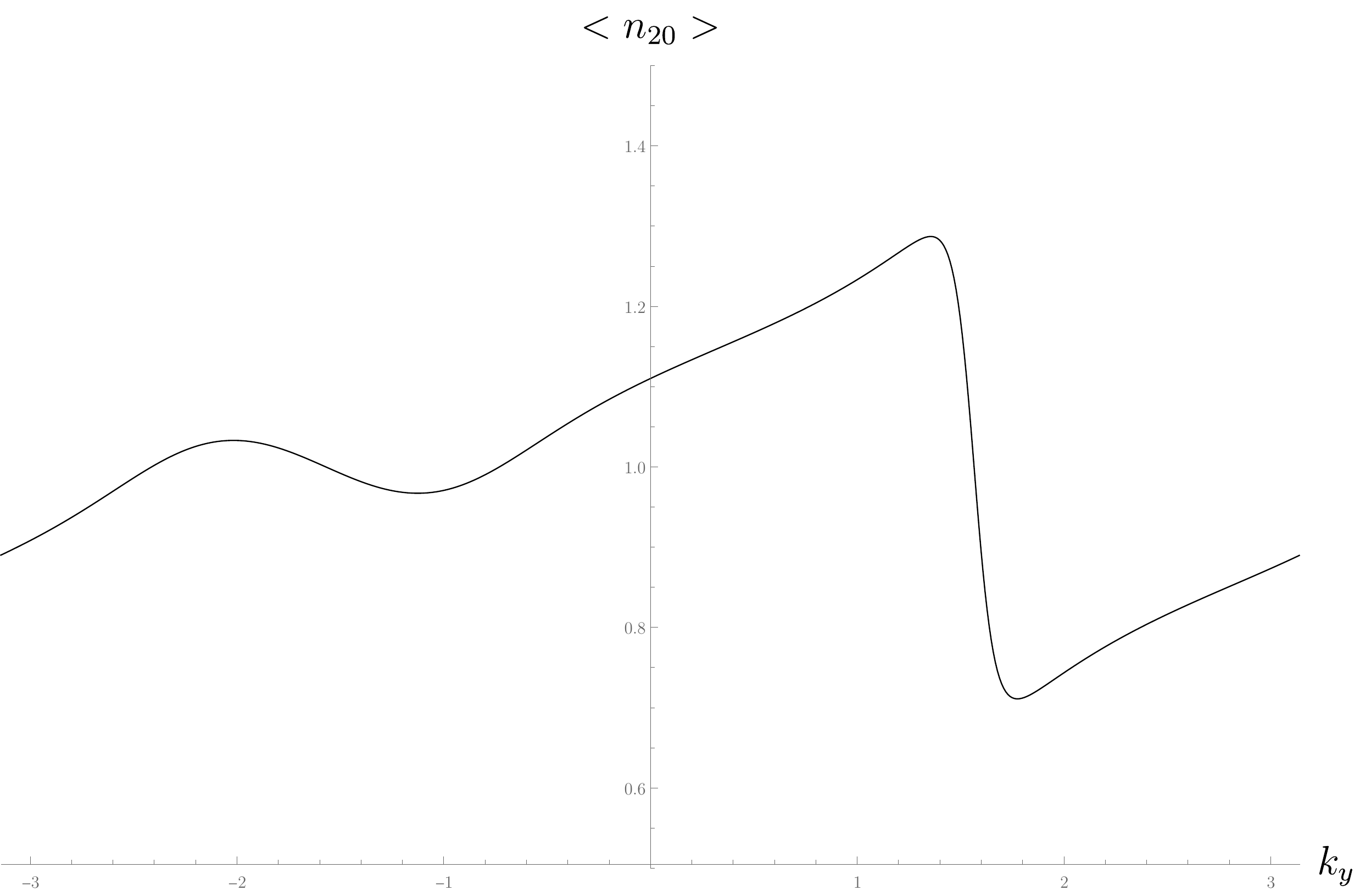}}

\subfloat[Occupation number in the bulk, at $T=10$.]{\includegraphics[scale=0.4]{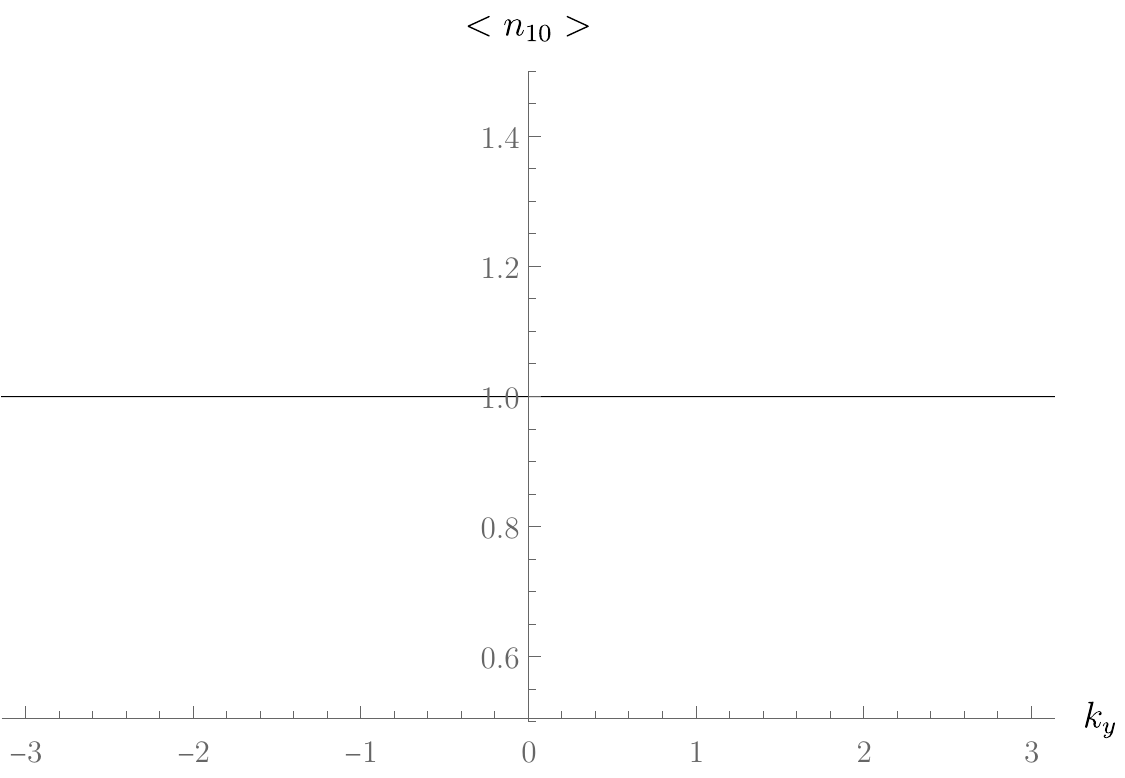}}
\qquad
\subfloat[Occupation number at the left edge, at $T=10$.]{\includegraphics[scale=0.4]{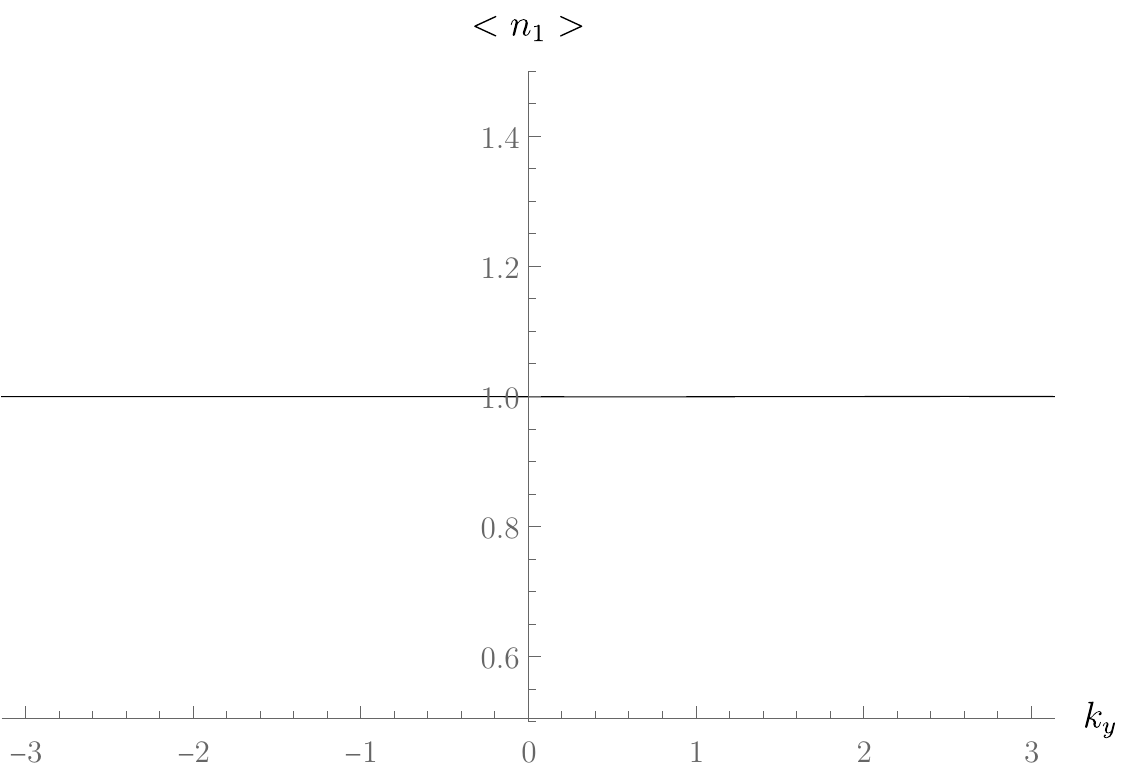}
}
\qquad
\subfloat[Occupation number at the right edge, at $T=10$.]{\includegraphics[scale=0.4]{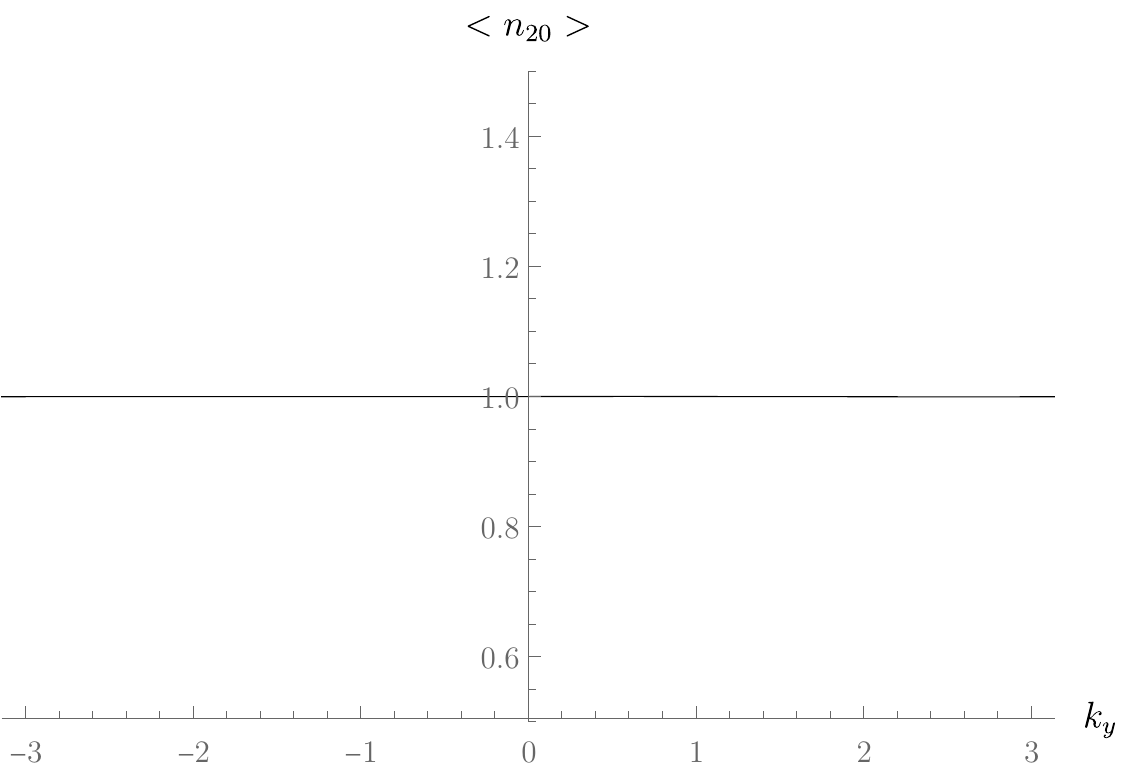}}

\centering
\caption{Expectation value of the occupation number at the bulk and boundary as a function of the quasi-momentum $k_{y}$ for the 2D topological insulator. The results are obtained for a chain of $20$ sites, $t_2=1.5$, a topologically non-trivial phase, $\mu=0.01$ and temperatures $T=1/10,\ 10$.}
\label{Fig: occupation numbers}
\end{figure}

In contrast, the spectrum of the topologically non-trivial system is composed of the previous two sets of bands and additional midgap states which are localized at the boundary of the system (c.f. Fig.~\ref{fig:TI spectrum non-trivial phase cylinder topology}). By adding a small negative chemical potential $\mu$ (smaller than the gap), since the edge states are localized and are midgap states, the expectation value of the particle number $\langle n_{i}(k_y) \rangle$ on the zero temperature limit of the density matrix will vary drastically at $k_y$ near the critical momentum where the gap closes and $i$ at the edge, i.e., $i=1$ and $i=N$. As the temperature raises, the particle numbers will saturate, since the density matrix tends to the totally mixed state. Indeed, this is what we can observe for the 2D TI in Fig.~\ref{Fig: occupation numbers}. A big variation in particle number occurs at the edges and in the vicinity of the critical momentum $k_y=\pi/2$, signalling the existence of edge states.

\newpage

\begin{figure}[h!]

\subfloat[$t_{2}=-2.5$]{\includegraphics[width=2in,height=1.5in]{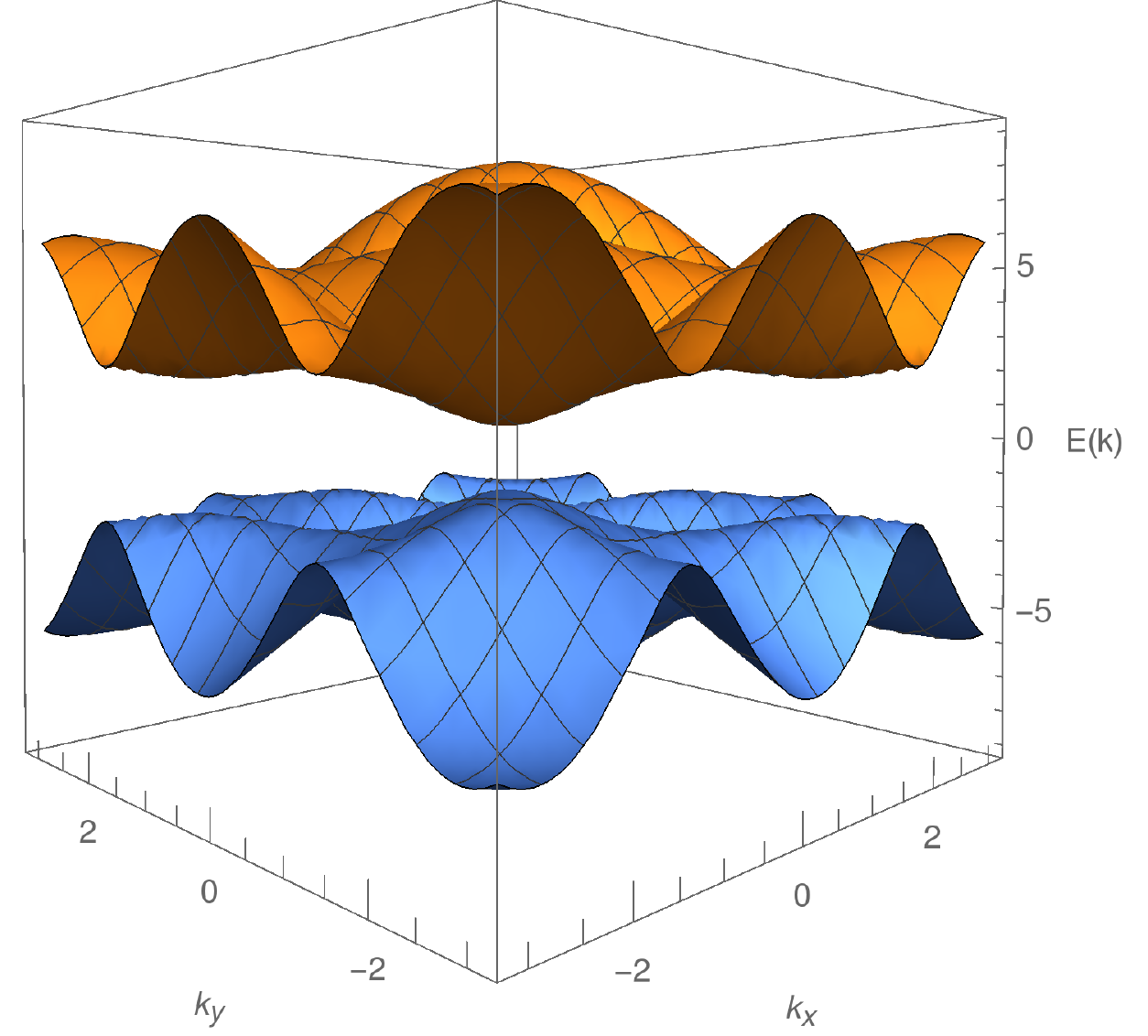}}
\subfloat[$t_{2}=-2.0$]{\includegraphics[width=2in,height=1.5in]{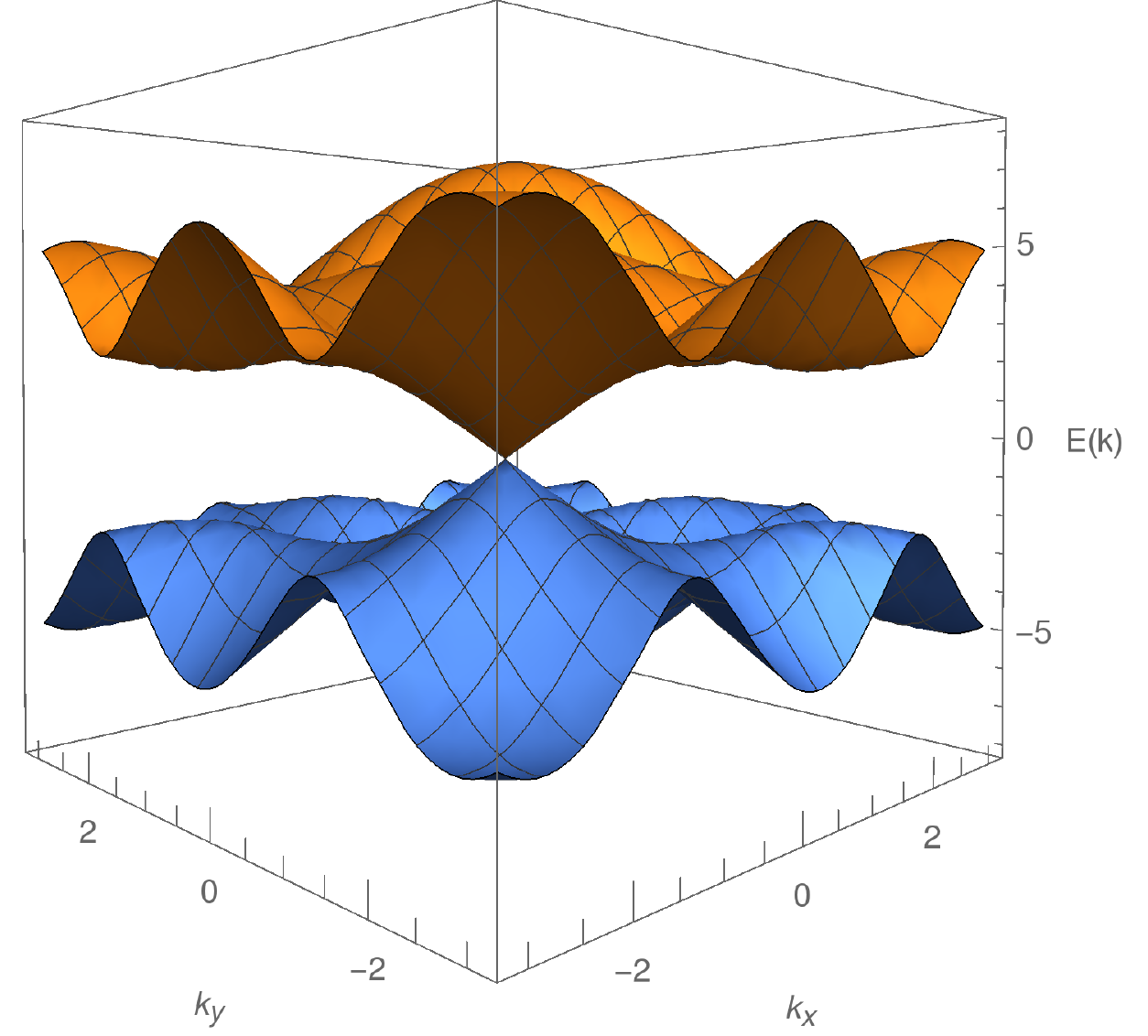}}
\subfloat[$t_{2}=-1.5$]{\includegraphics[width=2in,height=1.5in]{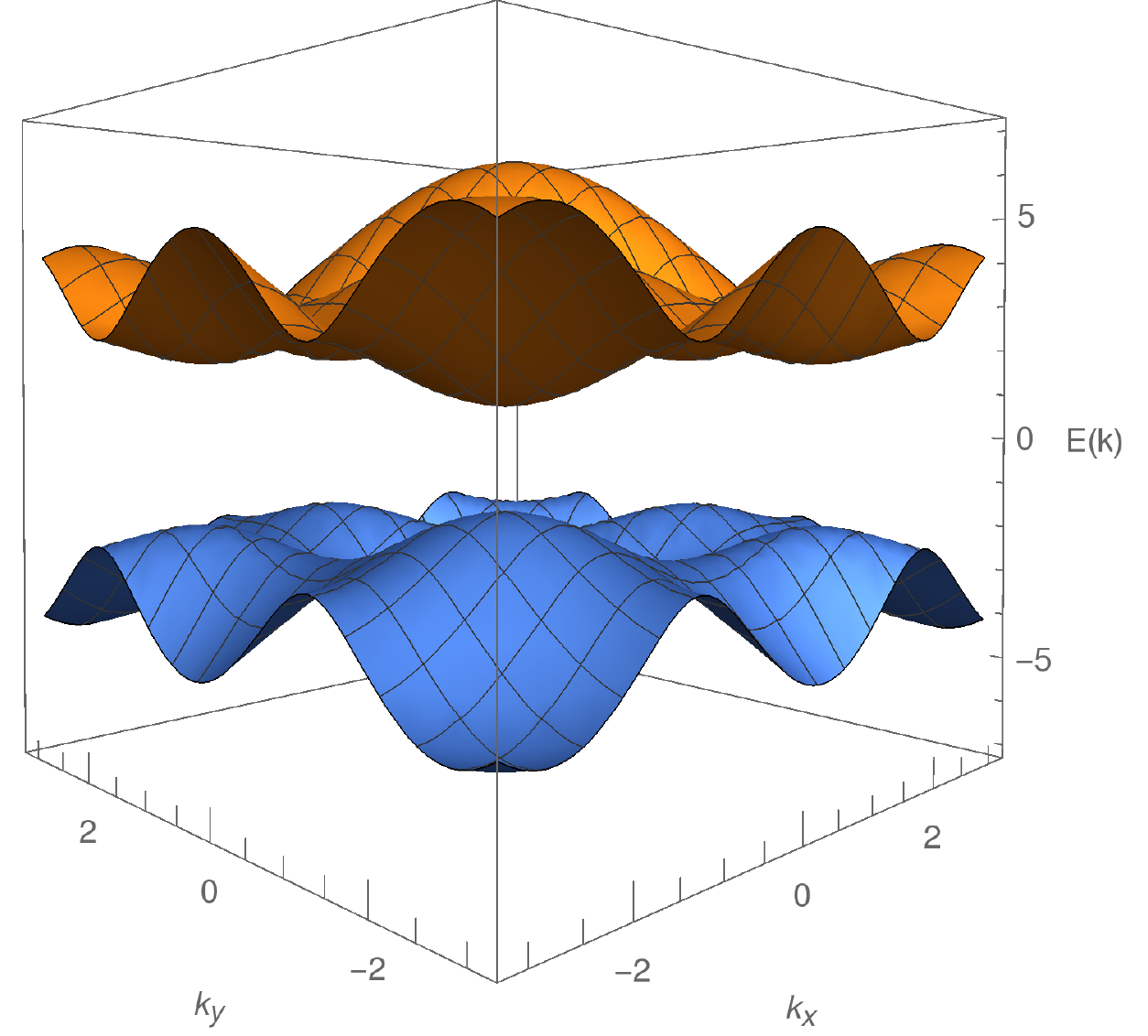}}

\subfloat[$t_{2}=0.0$]{\includegraphics[width=2in,height=1.5in]{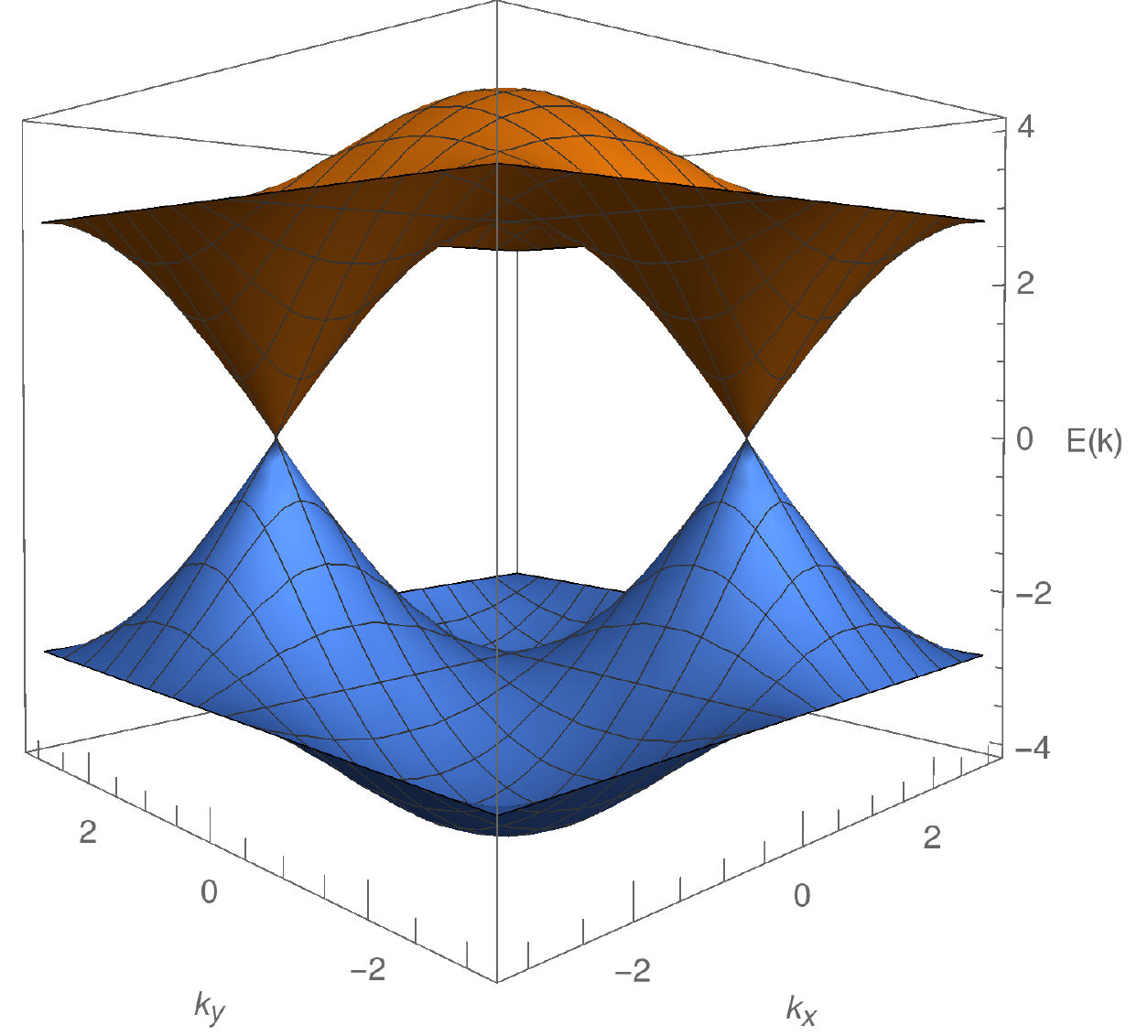}}
\subfloat[$t_{2}=1.5$]{\includegraphics[width=2in,height=1.5in]{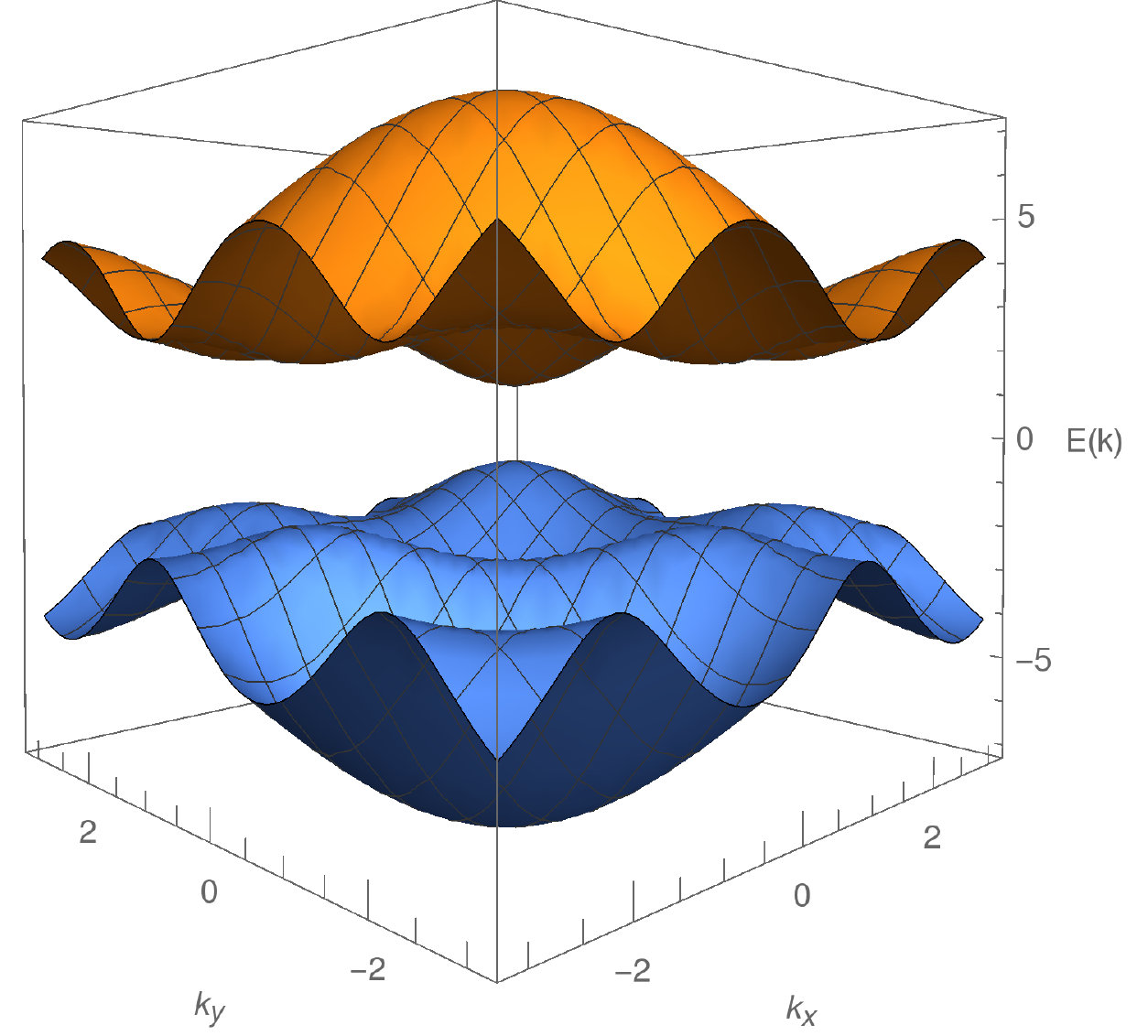}}
\subfloat[$t_{2}=2.0$]{\includegraphics[width=2in,height=1.5in]{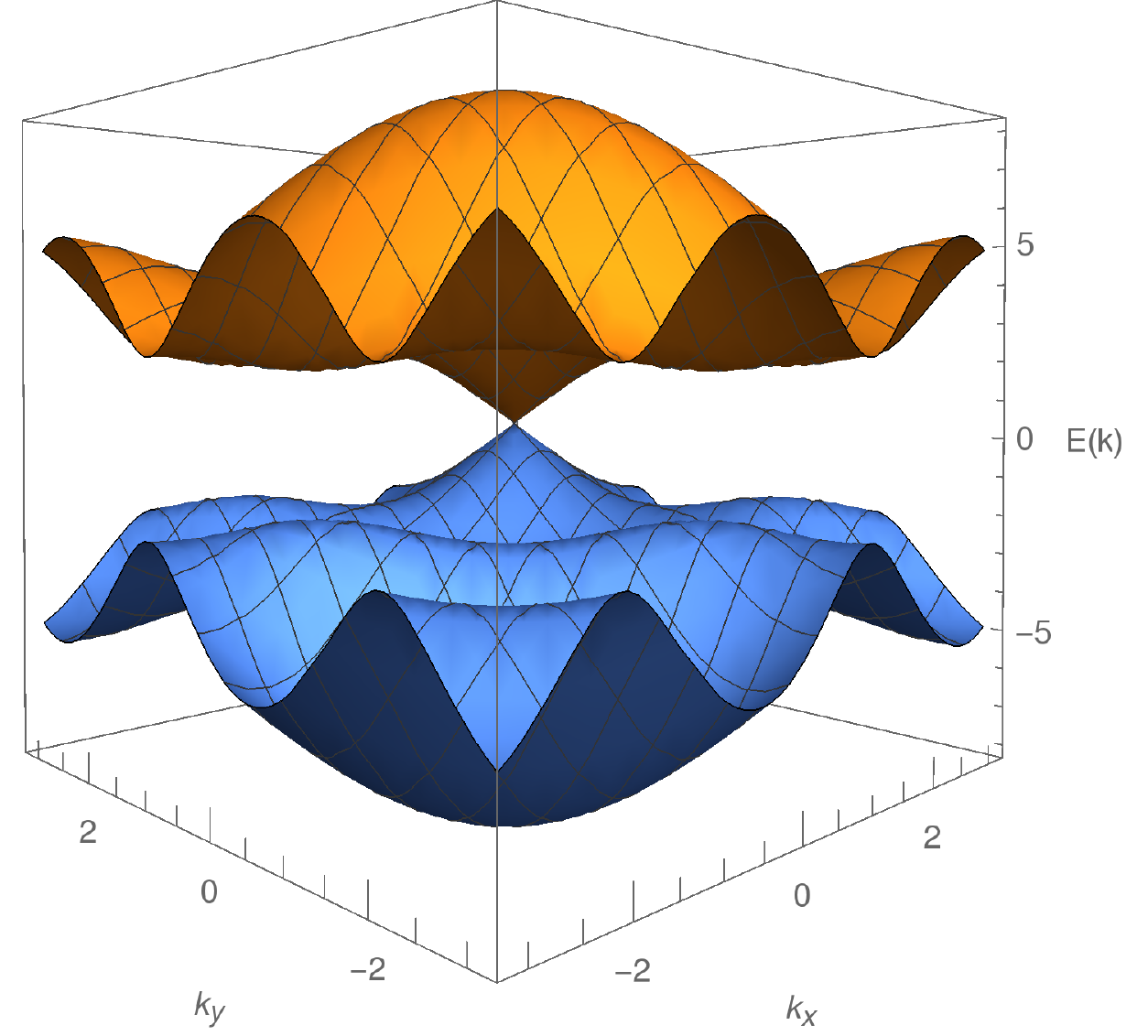}}

\centering
\caption{The bulk dispersion relation at different values of $t_{2}$ for the 2D topological insulator. In (a), (c) and (e), we show the cases of the system with a bulk gap (insulating phase), while in (b),(d) and (f) the system has no bulk gap.}
\label{fig:The-bulk-dispersion_For_TI-1}
\end{figure}

To find the bulk spectrum we used Eq.~\eqref{eq:H(kx, ky)}. We can also write Eq.~\eqref{eq:H(kx, ky)} as $H(k_{x},k_{y})=\vec{d}(k_x,k_y)\cdot \vec{\sigma}$, where $\vec{d}=(d_x,d_y,d_z)=(2t_{1}\cos(k_{x}),2t_{1}\cos(k_{y}),2t_{2}\cos(k_{x}+k_{y})+2t_{3}[\sin(k_{x})+\sin(k_{y})])$ and $\vec{\sigma}=(\sigma_{x},\sigma_{y},\sigma_{z})$. The bulk spectrum consists of two bands given by $E(k_{x},k_{y})=\pm|\mathbf{d}|$. For fixed $t_{1}=t_{3}=1$, the spectrum is plotted for various values of $t_{2}$, as shown in FIG.~\ref{fig:The-bulk-dispersion_For_TI-1}. We can see that the energy gap closes at the values $t_{2}=-2.0,\ 0,\ +2.0$. This can be evaluated by noticing that the closing of the gap is equivalent to $\vec{d}=0$. To have this, we need $d_{x}=0$ and $d_{y}=0$, which implies $\cos(k_x)=\cos(k_y)=0$. By inspection of $d_z$, we find the following four different points in Brillouin zone where gap closes. For $t_{2}=-2$, the gap closes at $k_{x}=k_{y}=-\frac{\pi}{2}$, whereas for $t_{2}=0$, the gap closes at two different points $k_{x}=\frac{\pi}{2},k_{y}=-\frac{\pi}{2}$ and $k_{x}=-\frac{\pi}{2},k_{y}=\frac{\pi}{2}$ and for $t_{2}=2$ the gap closes at $k_{x}=k_{y}=\frac{\pi}{2}$ as shown in FIG.~\ref{fig:The-bulk-dispersion_For_TI-1}. For $t_{2}\neq-2.0,\ 0,\ +2.0$ the system is gapped.

\section{Conclusions and future work}

In this paper, we applied the fidelity and Uhlmann connection analysis of phase transitions~\cite{mer:vla:pau:vie:17} to 2D free-fermion topological superconductors and insulators. First, we showed that the fidelity and the quantity $\Delta$ associated to the Uhlmann factor are detecting zero-temperature topological quantum phase transitions, as they assume non-trivial values around the gap-closing points. At finite temperatures, we probed both quantities with respect to small variations of the parameter of the Hamiltonian and the temperature. This two-fold analysis confirmed the fact that the two quantities behave in a different way, with respect to changes of the spectrum and the eigenbasis. Specifically, when we kept the temperature constant, and varied the parameter of the Hamiltonian, which affects both the spectrum and the eigenbasis of the state, the fidelity and $\Delta$ were signalling the topological phase transitions. On the other hand, when we varied the temperature alone, which only affects the spectrum, the fidelity was still detecting the topological phase transitions, while $\Delta$ stayed trivial, since it is only sensitive to changes of the eigenbasis of the state. 

Furthermore, we observed that as we increase the temperature, the non-triviality of fidelity and $\Delta$ around the gap-closing points at zero temperature is gradually smeared out and both quantities become trivial. This indicates that there are no finite-temperature phase transitions and the topological features of the systems at $T=0$ are gradually washed away. Furthermore, we analyzed the non-commutativity of the thermodynamic and zero temperature limits of the fidelity susceptibility and gave an analytical proof of the absence of finite temperature phase transitions in the class of models considered. This proof also complements the numerical results found in our previous work of Ref.~\cite{mer:vla:pau:vie:17}. Moreover, we analysed the critical behaviour at zero temperature by looking at the critical exponents of the fidelity susceptibility in the vicinity of the critical points.

We also performed a detailed study of the edge states of the systems at finite temperature. Our results show that as the temperature is increased the prominent -- at $T=0$ -- edge states start mixing with the bulk states, confirming the conclusion of the fidelity and $\Delta$ analysis, i.e., the gradual smearing of the topological features at finite temperatures (see also a related recent study~\cite{riv:viy:del:13}).

Finally, we would like to point out some directions of future work. First, one may consider the case of Hamiltonians with interactions or systems of bosons. A slightly different direction would be to use the same approach in the case of open systems, in which also the system-bath interaction should be taken into account. Moreover, the chiral p-wave superconductor that we considered in our analysis is shown to host vortices that obey non-abelian statistics, which are the main constituent of proposals for fault-tolerant quantum computation~\cite{kit:03,nay:sim:ste:fre:das:08} and the design of stable quantum memories~\cite{bro:los:pac:sel:woo:16}, due to their robustness against perturbations of the Hamiltonian. While the robustness of these modes has already been considered in the literature, it would be relevant to perform a more detailed quantitative analysis of their behaviour with respect to temperature, using our approach for the study of the edge states. Additionally, in this case the zero energy Majorana bound states appearing due to the vortices should be relevant in the behavior of the Uhlmann factor and deserve a separate and thorough study.

\acknowledgements{S. T. A., B. M. and C. V. acknowledge the support from DP-PMI and FCT (Portugal) through the grants PD/ BD/113651/2015, SFRH/BD/52244/2013 and PD/BD/52652/2014, respectively. BM thanks the support from Funda\c{c}\~{a}o para a Ci\^{e}ncia e Tecnologia (Portugal) namely through programmes PTDC/POPH/POCH and projects UID/EEA/50008/2013, IT/QuSim, IT/QuNet, ProQuNet, partially funded by EU FEDER, from the
EU FP7 project PAPETS (GA 323901) and from the JTF project NQuN (ID 60478). N. P. acknowledges the IT project QbigD funded by FCT PEst-OE/EEI/LA0008/2013, UID/EEA/50008/2013 and the Confident project  PTDC/EEI-CTP/4503/2014. Support from FCT (Portugal) through Grant UID/CTM/04540/2013 is also acknowledged.}

\appendix

\section{Analytic Expression for Fidelity}
\label{sec:Analytic-Expression-for-fidelity}
Let $\rho=\exp(-\beta H)$ and $\rho'=\exp(-\beta'H')$
be two unnormalized thermal states. In order to calculate an analytic expression for the fidelity we should first calculate the expression for $C$ through the following relation 
\begin{align}
e^{A}e^{B}e^{A} & =e^{C} \Leftrightarrow e^{A}e^{B}=e^{C}e^{-A},
\label{eq:UnknownC-1}
\end{align}
where $A=-\beta H$ and $B=-\beta'H'$.

Since the Hamiltonians $H$ and $H'$ are of the
form $H=\vec{d}\cdot\vec{\sigma}$, we can write 
\begin{align}
e^{A} & =a_{0}+\vec{a}\cdot\vec{\sigma},\nonumber \\
e^{B} & =b_{0}+\vec{b}\cdot\vec{\sigma},\label{eq:expInFormSigma-1}\\
e^{C} & =c_{0}+\vec{c}\cdot\vec{\sigma},\nonumber 
\end{align}
where the coefficients are real, since the Pauli matrices are Hermitian. Therefore, 
\begin{align*}
\det(e^{A}) & =a_{0}^{2}+\vec{a}^{2}=1 \\
\det(e^{B}) & =b_{0}^{2}+\vec{b}^{2}=1\\
\det(e^{C}) & =c_{0}^{2}+\vec{c}^{2}=1. 
\end{align*}

Using Eq.~(\ref{eq:expInFormSigma-1}) in Eq.~(\ref{eq:UnknownC-1}), and simplifying we get
\begin{align}
a_{0}b_{0}+\vec{a}.\vec{b}-a_{0}c_{0}+\vec{a}.\vec{c} & =0\nonumber \\
a_{0}\vec{b}+b_{0}\vec{a}+c_{0}\vec{a}-a_{0}\vec{c} & =0\label{eq:EqforCandCnot-1}\\
\vec{a}\times(\vec{b}-\vec{c}) & =0.\nonumber 
\end{align}
The last equation implies $\vec{c}=\vec{b}+\lambda\vec{a}$, with $\lambda\in \mathbb{R}$. By making this substitution in (\ref{eq:EqforCandCnot-1}), we obtain
\begin{equation*}
\begin{array}{lcl}
a_{0}c_{0}-\lambda\vec{a}^{2} & = & a_{0}b_{0}+2\vec{a}\cdot\vec{b} \\
(\lambda a_{0}-c_{0})\vec{a} & = & b_{0}\vec{a}.
\end{array}
\end{equation*}
In what follows we will determine the values of $c_{0}$ and $\lambda$:
\begin{align*}
\left[\begin{array}{cc}
a_{0} & -\vec{a}^{2}\\
-1 & a_{0}
\end{array}\right]\left[\begin{array}{c}
c_{0}\\
\lambda
\end{array}\right] & =\left[\begin{array}{c}
a_{0}b_{0}+2\vec{a}\cdot\vec{b}\\
b_{0}
\end{array}\right]
\end{align*}
\begin{align*}
\left[\begin{array}{c}
c_{0}\\
\lambda
\end{array}\right] & =\frac{1}{a_{0}^{2}-\vec{a}^{2}}\left[\begin{array}{cc}
a_{0} & \vec{a}^{2}\\
1 & a_{0}
\end{array}\right]\left[\begin{array}{c}
a_{0}b_{0}+2\vec{a}\cdot\vec{b}\\
b_{0}
\end{array}\right]\end{align*}

\begin{align*}
\left[\begin{array}{c}
c_{0}\\
\lambda
\end{array}\right] & =\left[\begin{array}{c}
(2a_{0}^{2}-1)b_{0}+2a_{0}\vec{a}\cdot\vec{b}\\
2(a_{0}b_{0}+\vec{a}\cdot\vec{b})
\end{array}\right].
\end{align*}
We proceed by writing $A=-\frac{\beta H}{2}:=-\frac{\epsilon}{2}\vec{x}\cdot\vec{\sigma}$ and $B=-\beta'H':=-\epsilon'\vec{y}\cdot\vec{\sigma}$ and we get
\begin{equation}
\begin{array}{lcl}
e^{A}=e^{-\frac{\epsilon}{2}\vec{x}\cdot\vec{\sigma}} & = &\cosh(\epsilon/2)-\sinh(\epsilon/2)\vec{x}\cdot\vec{\sigma} \\ [2mm]
e^{B}=e^{-\epsilon'\vec{y}\cdot\vec{\sigma}} & = &\cosh(\epsilon')-\sinh(\epsilon')\vec{y}\cdot\vec{\sigma}.
\end{array}
\label{eq:ActualHamiltonianForm-1}
\end{equation}
From (\ref{eq:expInFormSigma-1}) and (\ref{eq:ActualHamiltonianForm-1}), we get 
\begin{align*}
a_{0}=\cosh(\epsilon/2); & \ \ \ \vec{a}=-\sinh(\epsilon/2)\vec{x} \\
b_{0}=\cosh(\epsilon'); & \ \ \ \vec{b}=-\sinh(\epsilon')\vec{y}.
\end{align*}
Thus,
\begin{align*}
c_{0}= & (2a_{0}^{2}-1)b_{0}+2a_{0}\vec{a}.\vec{b}\\
= & (2\cosh^{2}(\epsilon/2)-1)\cosh(\epsilon')+2\cosh(\epsilon/2)\sinh(\epsilon/2)\sinh(\epsilon')\vec{x}\cdot\vec{y}\\
= & \cosh(\epsilon)\cosh(\epsilon')+\sinh(\epsilon)\sinh(\epsilon')\vec{x}\cdot\vec{y}.
\end{align*}

Since the product of Pauli matrices is again a Pauli matrix, we can write $C=\xi\vec{z}\cdot\vec{\sigma}$,
\begin{align*}
e^{C}=e^{\xi\vec{z}\cdot\vec{\sigma}} & =\cosh(\xi)-\sinh(\xi)\vec{z}\cdot\vec{\sigma}\\
\cosh(\xi)=c_{0} & =\cosh(\epsilon)\cosh(\epsilon')+\sinh(\epsilon)\sinh(\epsilon')\vec{x}\cdot\vec{y}.
\end{align*}
The unnormalized expression for the fidelity is 
\begin{align*}
F_{\text{un}} & =\tr\sqrt{e^{-\frac{\epsilon}{2}\vec{x}.\vec{\sigma}}e^{-\epsilon'\vec{y}\cdot\vec{\sigma}}e^{-\frac{\epsilon}{2}\vec{x}\cdot\vec{\sigma}}}=\tr (e^{C/2})\\
 & =\tr(\cosh(\xi/2)-\sinh(\xi/2)\vec{z}\cdot\vec{\sigma})=2\cosh(\xi/2).
\end{align*}
By substituting $\cosh(\xi/2)$, we get
\begin{align*}F_{\text{un}}= & \sqrt{2(1+\cosh(\epsilon)\cosh(\epsilon')+\sinh(\epsilon)\sinh(\epsilon')\vec{x}\cdot\vec{y})}.
\end{align*}
For $\epsilon=\beta E_{\mathbf{k}}/2$, 
$\vec{x}=\vec{n}_{\mathbf{k}}$, $\epsilon'=\beta'E'_{\mathbf{k}}/2$ and $\vec{y}=\vec{n}'_{\mathbf{k}}$ with $\mathbf{k}=(k_{x},k_{y})$, the above equation becomes

\begin{align*}
F_{\text{un}}= \tr e^{C_\mathbf{k}/2}= \sqrt{2(1+\cosh(E_\mathbf{k}/2T)\cosh(E'_\mathbf{k}/2T')+\sinh(E_\mathbf{k}/2T)\sinh(E'_{\mathbf{k}}/2T')\vec{n}_\mathbf{k}\cdot\vec{n'}_\mathbf{k})}.
\end{align*}

To compute the final expression for the fidelity, we will use 
\begin{align*}
\tr(e^{-\beta\mathfrak{\mathcal{H}}})=\tr(e^{-\beta\Psi^{\dagger}H\Psi}) = \text{det}(I+e^{-\beta H})=1+\tr(e^{-\beta H})+\text{det}(e^{-\beta H}),
\end{align*}
and we obtain

\begin{align*}
F(\rho,\rho') & =\prod_{\mathbf{k}\epsilon\mathcal{B}}\frac{\tr(e^{-\Psi^{\dagger} C_{\mathbf{k}} \Psi/2})}{\sqrt{\tr(e^{-\beta\mathcal{H}_{\mathbf{k}}})\tr(e^{-\beta'\mathcal{H'}_{\mathbf{k}}})}}=\prod_{\mathbf{k}\epsilon\mathcal{B}}\frac{\tr(I+e^{-C_{\mathbf{k}}/2})}{\sqrt{\tr(I+e^{-\beta H_{\mathbf{k}}})\tr(I+e^{-\beta' H'_{\mathbf{k}}})}} \\ 
& =\prod_{\mathbf{k}\epsilon\mathcal{B}}\frac{2+\sqrt{2(1+\cosh(E_{\mathbf{k}}/2T)\cosh(E'_{\mathbf{k}}/2T')+\sinh(E_{\mathbf{k}}/2T)\sinh(E'_{\mathbf{k}}/2T')\vec{n}_{\mathbf{k}}.\vec{n'}_{\mathbf{k}})}}{\sqrt{(2+2\cosh(E_{\mathbf{k}}/2T))(2+2\cosh(E'_{\mathbf{k}}/2T'))}}.
\end{align*}

Similarly, we can calculate $\tr\sqrt{\rho}\sqrt{\rho'}$ and determine $\Delta(\rho,\rho')$

\begin{align*}
\tr\sqrt{\rho}\sqrt{\rho'}=\prod_{\mathbf{k}\epsilon\mathcal{B}}\frac{2+2(\cosh(E_{\mathbf{k}}/4T)\cosh(E'_{\mathbf{k}}/4T')+\sinh(E_{\mathbf{k}}/4T)\sinh(E'_{\mathbf{k}}/4T')\vec{n}_{\mathbf{k}}\cdot\vec{n'}_{\mathbf{k}})}{\sqrt{(2+2\cosh(E_{\mathbf{k}}/2T))(2+2\cosh(E'_{\mathbf{k}}/2T'))}}.
\end{align*}

\section{Hamiltonian transformations \label{sec:Hamiltonian-transformations}}

Let $A=t_{1}\sigma_{x}+it_{3}\sigma_{z}$, $B=t_{1}\sigma_{y}+it_{3}\sigma_{z}$ and $C=t_{2}\sigma_{z}$. Then,
\begin{eqnarray}
\sum_{j}c_{i+1,j}^\dagger Ac_{i,j} & = & \sum_{k_{y}}c_{i+1,k_{y}}^\dagger Ac_{i,k_{y}} \label{eq:A-1} \\
\sum_{j}c_{i,j+1}^\dagger Bc_{i,j} & = &\sum_{k_{y}}\exp(ik_{y})c_{i,k_{y}}^\dagger Bc_{i,k_{y}}\label{eq:B-1} \\
\sum_{j}c_{i+1,j+1}^\dagger Cc_{i,j} & = &\sum_{k_{y}}\exp(ik_{y})c_{i+1,k_{y}}^\dagger Cc_{i,k_{y}}\label{eq:C-1}.
\end{eqnarray}
By taking the Hermitian conjugate of the above three equations, we get
\begin{eqnarray}
\sum_{j}c_{i,j}^\dagger A^{\dagger}c_{i+1,j} & = &\sum_{k_{y}}c_{i,k_{y}}^\dagger A^{\dagger}c_{i+1,k_{y}}\label{eq:A*-1} \\
\sum_{j}c_{i,j}^\dagger B^{\dagger}c_{i,j+1} & = &\sum_{k_{y}}\!\exp(-ik_{y})c_{i,k_{y}}^\dagger \! B^{\dagger}c_{i,k_{y}}\label{eq:B*-1} \\
\sum_{j}c_{i,j}^\dagger C^{\dagger}c_{i+1,j+1} & = & \sum_{k_{y}}\!\exp(-ik_{y})c_{i,k_{y}}^\dagger \! C^{\dagger}c_{i+1,k_{y}}\label{eq:C*-1}.
\end{eqnarray}
Summing Eq.~(\ref{eq:A-1}) and~(\ref{eq:C-1}), we get
\begin{align*}
 & \sum_{k_{y}}c_{i+1,k_{y}}^\dagger Ac_{i,k_{y}}+\sum_{k_{y}}\exp(ik_{y})c_{i+1,k_{y}}^\dagger Cc_{i,k_{y}} \\
 & =\sum_{k_{y}}c_{i+1,k_{y}}^\dagger (t_{1}\sigma_{x}+it_{3}\sigma_{z}+t_{2}\exp(ik_{y})\sigma_{z})c_{i,k_{y}}.
\end{align*}
Accordingly, from Eq.~(\ref{eq:B-1}) and~(\ref{eq:B*-1}) 
\begin{align*}
 & \!\!\!\!\!\!\!\!\!\! \sum_{k_{y}}\exp(ik_{y})c_{i,k_{y}}^\dagger Bc_{i,k_{y}}+\sum_{k_{y}}\exp(-ik_{y})c_{i,k_{y}}^\dagger B^{\dagger}c_{i,k_{y}} \\
 & =\sum_{k_{y}}c_{i,k_{y}}^\dagger (2\cos(k_{y})t_{1}\sigma_{y}-2\sin(k_{y})t_{3}\sigma_{z})c_{i,k_{y}},
\end{align*}
and finally from Eq.~(\ref{eq:A*-1}) and (\ref{eq:C*-1})
\begin{align*}
 & \sum_{k_{y}}c_{i,k_{y}}^\dagger A^{\dagger}c_{i+1,k_{y}}+\sum_{k_{y}}\exp(-ik_{y})c_{i,k_{y}}^\dagger C^{\dagger}c_{i+1,k_{y}} \\
 & =\sum_{k_{y}}c_{i,k_{y}}^\dagger (t_{1}\sigma_{x}-it_{3}\sigma_{z}+t_{2}\exp(-ik_{y})\sigma_{z})c_{i+1,k_{y}}.
\end{align*}
Thus the total Hamiltonian of the topological insulator with transitional invariance in the $y$-axis becomes
\begin{align*}
H = & \sum_{k_{y}}\sum_{i}(c_{i+1,k_{y}}^\dagger (t_{1}\sigma_{x}+it_{3}\sigma_{z}+t_{2}\exp(ik_{y})\sigma_{z})c_{i,k_{y}}\\
 & +c_{i,k_{y}}^\dagger (2\cos(k_{y})t_{1}\sigma_{y}-2\sin(k_{y})t_{3}\sigma_{z})c_{i,k_{y}}\\
 & +c_{i,k_{y}}^\dagger (t_{1}\sigma_{x}-it_{3}\sigma_{z}+t_{2}\exp(-ik_{y})\sigma_{z})c_{i+1,k_{y}}),
 \end{align*}
and for $H(k_{y})$ we have
\begin{align*}
\!H(k_{y}) \! = \! & \sum_{i,j}[c_{i,k_{y}}^\dagger \! (t_{1}\sigma_{x} \! + \! it_{3}\sigma_{z} \! + \! t_{2}\exp(ik_{y})\sigma_{z})\delta_{i,j+1}c_{j,k_{y}}\\
& \! +c_{i,k_{y}}^\dagger \! (2\cos(k_{y})t_{1}\sigma_{y} \! - \! 2\sin(k_{y})t_{3}\sigma_{z})\delta_{i,j}c_{j,k_{y}}\\
& \! +c_{i,k_{y}}^\dagger \! (t_{1}\sigma_{x} \! - \! it_{3}\sigma_{z} \! + \! t_{2}\exp(-ik_{y})\sigma_{z})\delta_{i+1,j}c_{j,k_{y}}],
\end{align*}
with the $(i,j)$ matrix elements of $H(k_{y})$ being 
\begin{eqnarray}
H_{i,j}(k_{y}) & = &(t_{1}\sigma_{x}+it_{3}\sigma_{z}+t_{2}\exp(ik_{y})\sigma_{z})\delta_{i,j+1} \nonumber \\
&& +(2\cos(k_{y})t_{1}\sigma_{y}-2\sin(k_{y})t_{3}\sigma_{z})\delta_{i,j}\label{eq:Hij(kx)-1} \\
&& +(t_{1}\sigma_{x}-it_{3}\sigma_{z}+t_{2}\exp(-ik_{y})\sigma_{z})\delta_{i+1,j}. \nonumber
\end{eqnarray}
Simplifying Eq.~(\ref{eq:Hij(kx)-1}), we obtain
\begin{align}
H_{i,j}(k_{y}) \!\! & =  \! t_{1}(\delta_{i,j+1}+\delta_{i+1,j})\sigma_{x}+2t_{1}\cos(k_{y})\delta_{i,j}\sigma_{y}\nonumber\\
&\ \  + \! \{i(t_{3} \!\! + \! t_{2}\sin(k_{y}))(\delta_{i,j+1} \!\! - \! \delta_{i+1,j}) \!\! - \! 2t_{3}\sin(k_{y})\delta_{i,j}\nonumber\\
&\ \  +  t_{2}\cos(k_{y})(\delta_{i,j+1}+\delta_{i+1,j})\}\sigma_{z},
\label{eq:H_ij(k_{y})}
\end{align}
or 
\begin{align}
H_{i,j}(k_{y}) =a\sigma_{x}+b\sigma_{y}
+c\sigma_{z},
\end{align}
with $a=t_{1}(\delta_{i,j+1}+\delta_{i+1,j})$, $b=2t_{1}\cos(k_{y})\delta_{i,j}$ and $c=i(t_{3}+t_{2}\sin(k_{y}))(\delta_{i,j+1}-\delta_{i+1,j})-2t_{3}\sin(k_{y})\delta_{i,j}+t_{2}\cos(k_{y})(\delta_{i,j+1}+\delta_{i+1,j})$.

If we also consider periodic boundary conditions along the $x$-axis, we will have translational invariance, and after applying the full Fourier transform, we get the following bulk Hamiltonian for the topological insulator in~2D:
\begin{align*}
H(\mathbf{k}) = 2t_{1}\cos(k_{x})\sigma_{x}+2t_{1}\cos(k_{y})\sigma_{y}+\{2t_{2}\cos(k_{x}+k_{y})+2t_{3}[\sin(k_{x})+\sin(k_{y})]\}\sigma_{z}. 
\end{align*}

\section{Derivation of the Bures Metric}
\label{Appendix C}
Consider a generic charge symmetric fermionic Gaussian state
\begin{align*}
\rho=\exp(X), \text{ with } X=-\psi^{\dagger}H\psi -\log Z,\ \log Z=\tr e^{-\psi^{\dagger}H\psi} \text{ and } H\in\mbox{Herm}(\mathbb{C}^N).
\end{align*}
Here $\psi$ and $\psi^{\dagger}$ are the usual arrays of fermion destruction and creation operators. We are interested in evaluating the pullback of the Bures metric by the map $\rho: \mbox{Herm}(\mathbb{C}^N)\ni H \mapsto e^{X}$. Now, the Bures metric is given by
\begin{align*}
ds^2=\tr\{\rho G^2\},
\end{align*}
with
\begin{align*}
d\rho = G\rho +\rho G.
\end{align*}
The last equation is equivalent to (for invertible states, which is the case at hand)
\begin{align*}
\rho^{-1}d\rho= \rho^{-1}G\rho +G = (1 + e^{-\mbox{ad}_X})(G),
\end{align*}
where $\mbox{ad}_X(\cdot):=[X,\cdot]$. Now,
\begin{align*}
d\rho =d e^{X}= e^{X}(e^{-X} d e^{X})=e^{X}\int_{0}^{1}ds\ e^{-s X} dX e^{sX},
\end{align*}
since
\begin{align*}
\frac{\partial}{\partial s}\big(e^{-sX}d(e^{sX})\big)=e^{-sX}dX e^{sX}.
\end{align*}
Therefore,
\begin{align*}
d\rho =e^{X} \int_{0}^{1}ds\ e^{-s\ \mbox{ad}_{X}}(dX)=-e^{X}(\mbox{ad}_X)^{-1} (e^{-\mbox{ad}_X} -1)(dX),
\end{align*}
and
\begin{align*}
\rho^{-1}d\rho= (\mbox{ad}_X)^{-1} (1-e^{-\mbox{ad}_X})(dX).
\end{align*}
Thus,
\begin{align*}
(1 + e^{-\mbox{ad}_X})(G)=(\mbox{ad}_X)^{-1} (1-e^{-\mbox{ad}_X})(dX)
\end{align*}
Formally,
\begin{align*}
G=\frac{1}{2}\frac{\tanh\big(\frac{{\mbox{ad}_X}}{2}\big)}{\frac{{\mbox{ad}_X}}{2}}(dX).
\end{align*}
Notice that
\begin{align*}
dX=-\psi^{\dagger}dH\psi -\frac{dZ}{Z}.
\end{align*}
Using the fact that any Hermitian matrix can be diagonalized by a unitary matrix, we can write,
\begin{align*}
H=SDS^{-1}, \text{ with } S\in\mbox{U}(N) \text { and } D=\mbox{diag}(\lambda_1,...,\lambda_N). 
\end{align*}
then,
\begin{align*}
dH=S([S^{-1}dS,D] +dD)S^{-1}.
\end{align*}
Then,
\begin{align*}
\psi^{\dagger} S =\widetilde{\psi}^{\dagger} \text{ and } S^{-1}\cdot \psi=\widetilde{\psi},
\end{align*}
satisfy the usual canonical anticommutation relations. Moreover,
\begin{align*}
&\mbox{ad}_{X}(\widetilde{\psi}^{\dagger}_i)=-\lambda_i\widetilde{\psi}^{\dagger}_i,\ i=1,...,N,\\
&\mbox{ad}_{X}(\widetilde{\psi}_i)=\lambda_i\widetilde{\psi}_i,\ i=1,...,N.
\end{align*}
We can then write,
\begin{align*}
dX=-\widetilde{\psi}^{\dagger}\big([S^{-1}dS,D]+dD\big)\widetilde{\psi}-\frac{dZ}{Z}.
\end{align*}
and,
\begin{align*}
G &=\frac{1}{2}\frac{\tanh\big(\frac{{\mbox{ad}_X}}{2}\big)}{\frac{{\mbox{ad}_X}}{2}}(dX)\\
&=-\frac{1}{2}\sum_{i,j=1}^{N}\widetilde{\psi}^{\dagger}_{i}\Big(-\big(S^{-1}dS\big)_{ij}\frac{\tanh\big(\frac{\lambda_i-\lambda_j}{2}\big)}{\frac{\lambda_i-\lambda_j}{2}}(\lambda_i-\lambda_j)+d\lambda_i\delta_{ij}\Big)\widetilde{\psi}_j-\frac{1}{2}\frac{dZ}{Z}\\
 &=\frac{1}{2}\frac{\tanh\big(\frac{{\mbox{ad}_X}}{2}\big)}{\frac{{\mbox{ad}_X}}{2}}(dX)=\sum_{i,j=1}^{N}\widetilde{\psi}^{\dagger}_{i}\Big(\big(S^{-1}dS\big)_{ij}\tanh\big(\frac{\lambda_i-\lambda_j}{2}\big)-\frac{d\lambda_i}{2}\delta_{ij}\Big)\widetilde{\psi}_j-\frac{1}{2}\frac{dZ}{Z},
\end{align*}
so, clearly, $G$ is of the form
\begin{align*}
G=\psi^{\dagger}K\psi +C,
\end{align*}
for an Hermitian matrix valued one-form $K=SBS^{-1}$ and one-form $C$,
\begin{align*}
& B_{ij}=\big(S^{-1}dS\big)_{ij}\tanh\big(\frac{\lambda_i-\lambda_j}{2}\big)-\frac{d\lambda_i}{2}\delta_{ij},\\
& C=-\frac{1}{2}\frac{dZ}{Z}.
\end{align*}
The one-form $S^{-1}dS$ is the Maurer-Cartan differential form. To evaluate the metric tensor, we just need to compute the trace,
\begin{align*}
ds^2=\tr\big(\rho G^2\big)=\langle G^2\rangle =\langle \psi^{\dagger} K\psi \psi^{\dagger} K\psi\rangle +2\langle \psi^{\dagger} K\psi\rangle C +C^2.
\end{align*}
To evaluate the last expression, consider the correlation function
\begin{align*}
g_{ji}:=\langle \psi^{\dagger}_i\psi_j\rangle.
\end{align*}
It is clear that for any single particle observable $A\in\text{Herm}(\mathbb{C}^n)$, we have
\begin{align*}
\langle \psi^{\dagger} A\psi\rangle =\tr \big( A g\big).
\end{align*}
Moreover, we will use Wick's theorem to evaluate expressions of the form
\begin{align*}
\langle \psi^{\dagger}_i\psi_j\psi^{\dagger}_k\psi_l\rangle =g_{ji}g_{lk}+g_{li}(\delta_{jk}-g_{jk}).
\end{align*}
Then, we can write,
\begin{align*}
\langle \psi^{\dagger} K\psi \psi^{\dagger} K\psi\rangle &=K_{ij}K_{kl}\langle \psi^{\dagger}_{i}\psi_{j}\psi^{\dagger}_k\psi_l\rangle=K_{ij}K_{kl}\big(g_{ji}g_{lk}+g_{li}(\delta_{jk}-g_{jk}\big)\\
&=\big(\tr\big(Kg\big)\big)^2+\tr\big(K^2g\big)-\tr\big(gKgK\big),
\end{align*}
and
\begin{align*}
\langle\psi^{\dagger}K\psi\rangle =\tr\big(Kg\big).
\end{align*}
Moreover, since $\tr(d\rho)=0$, we have,
\begin{align*}
0=2\tr\big(\rho G\big) =2\left(\langle\psi^{\dagger}K\psi\rangle +C\right),
\end{align*}
hence
\begin{align*}
C=-\langle \psi^{\dagger} K \psi\rangle=-\tr \big(Kg\big)=-\frac{1}{2}\frac{dZ}{Z}.
\end{align*}
Finally, we can write,
\begin{align*}
ds^2&=\big(\tr\big(Kg\big)\big)^2+\tr\big(K^2g\big)-\tr\big(gKgK\big) -2\big(\tr\big(Kg\big)\big)^2 +\big(\tr\big(Kg\big)\big)^2\\
&=\tr\big(K^2g\big)-\tr\big(gKgK\big).
\end{align*}
Also as
\begin{align*}
ds^2=\langle \psi^{\dagger} K\psi \psi^{\dagger} K\psi\rangle +2\langle \psi^{\dagger} K\psi\rangle C +C^2 =\langle \big(\psi^{\dagger} K\psi\big)^2\rangle -\langle \psi^{\dagger} K\psi\rangle^2 .
\end{align*}
Next, notice that
\begin{align*}
\langle \widetilde{\psi}^{\dagger}_j\widetilde{\psi}_i\rangle=\delta_{ij}\frac{1}{e^{-\lambda_i} +1}\equiv \delta_{ij}n_i=S_{kj}\overline{S}_{li}\langle \psi_{k}^{\dagger}\psi_l\rangle=S^{-1}_{il}g_{lk}S_{kj},
\end{align*}
and hence,
\begin{align*}
g=(1+e^{-H})^{-1}.
\end{align*}
Now, defining $A_{ij}=(S^{-1}dS)_{ij}=-\overline{A}_{ji}$ and $t_{ij}=-t_{ji}=\tanh\big((\lambda_i-\lambda_j)/2\big)$, we have
\begin{align*}
\tr\big(K^2g\big)&=B_{ij}B_{jl}n_{l}\delta_{li}=n_{i}B_{ij}B_{ji}=n_{i}\Big[A_{ij}t_{ij}-\frac{d\lambda_i}{2}\delta_{ij}\Big]\Big[A_{ji}t_{ji}-\frac{d\lambda_j}{2}\delta_{ji}\Big]\\
&=n_i\big(|A_{ij}|^2t_{ij}^2 +\frac{1}{4}d\lambda_i^2\big),
\end{align*}
and
\begin{align*}
\tr\big(gKgK\big)&=n_{i}\delta_{ij}B_{jk}n_{k}\delta_{kl}B_{li}=n_{i}n_{j}B_{ij}B_{ji}\\
&=n_{i}n_{j}\Big[A_{ij}t_{ij}-\frac{d\lambda_i}{2}\delta_{ij}\Big]\Big[A_{ji}t_{ji}-\frac{d\lambda_j}{2}\delta_{ji}\Big]\\
&=n_in_j\big(|A_{ij}|^2t_{ij}^2 +\frac{1}{4}d\lambda_i^2\delta_{ij}\big),
\end{align*}
hence,
\begin{align*}
ds^2&=\sum_{i\neq j}n_i(1-n_j)|A_{ij}|^2t_{ij}^2 +\frac{1}{4}\sum_{i=1}^{N}n_i(1-n_i)d\lambda_i^2\\
&=\sum_{i\neq j}n_i(1-n_j)|(S^{-1} dS)_{ij}|^2\tanh^2\big(\frac{\lambda_i-\lambda_j}{2}\big) +\frac{1}{4}\sum_{i=1}^{N}n_i(1-n_i)d\lambda_i^2.
\end{align*}
The second term is the classical one, i.e., the classical Fisher information associated with the probability distribution,
\begin{align*}
p_{I}:=p_{i_1,...,i_p}=\frac{e^{-(\lambda_{i_1}+...+\lambda_{i_p})}}{Z},\ 1< i_1<...<i_p<N.
\end{align*}
Since,
\begin{align*}
dp_{I}=-\big[(d\lambda_{i_1}+...+d\lambda_{i_p})+d\log(Z)\big]p_{I},
\end{align*}
we have,
\begin{align*}
&\frac{1}{4}\sum_{I}\frac{dp_{I}^2}{p_I}\\
&=\frac{1}{4}\sum_{I}p_{I}\big[(d\lambda_{i_1}+...+d\lambda_{i_p})^2 +2(d\lambda_{i_1}+...+d\lambda_{i_p})d\log(Z) +d\log(Z)^2\big]\\
&=\frac{1}{4}\left[\langle \widetilde{\psi}^{\dagger} dD^2 \widetilde{\psi}\rangle +2\langle \widetilde{\psi}^{\dagger} dD  \widetilde{\psi}\rangle d\log(Z) +d\log(Z)^2\right]\\
&=\frac{1}{4}\left[\tr\big(g(SdDS^{-1})^2\big)+2\tr\big(gSdDS^{-1})d\log(Z)+d\log(Z)^2\right]
\end{align*}
But,
\begin{align*}
\frac{1}{2}d\log(Z)=\tr \big(gK\big)=n_iB_{ii}=-\frac{1}{2}\sum_{i=1}^{N}n_{i}d\lambda_i=-\frac{1}{2}\tr\big(gSdDS^{-1}\big),
\end{align*}
therefore,
\begin{align*}
\frac{1}{4}\sum_{I}\frac{dp_{I}^2}{p_I}=\frac{1}{4}\left[\tr\big(g(SdDS^{-1})^2\big)-\big(\tr\big(gSdDS^{-1}\big)\big)^2\right]=\frac{1}{4}\sum_{i=1}^N n_i(1-n_i)d\lambda_i^2.
\end{align*}
Also notice that this corresponds to evaluating the metric when $S^{-1}dS\equiv 0$. The non-classical part is therefore given by,
\begin{align*}
\sum_{i\neq j}n_i(1-n_j)|(S^{-1} dS)_{ij}|^2\tanh^2\big(\frac{\lambda_i-\lambda_j}{2}\big).
\end{align*}
\subsection{Two-level case}
We now suppose $N=2$ and without loss of generality we can write,
\begin{align*}
H=x^{\mu}\sigma_{\mu}=(U\sigma_3 U^{-1})r, \text{ with } r=\sqrt{x^{\mu}x^{\mu}}.
\end{align*}
The eigenstates of $H$ are coherent states of $\mbox{SU}(2)$, namely if we take $n^{\mu}=x^{\mu}/r$ and $z=(n^{1}+in^2)/(1+n^3)$ is the stereographic projection with respect to the south pole of the sphere, we can write, 
\begin{align*}
\ket{z}=\frac{1}{(1+|z|^2)^{1/2}}\left(\begin{array}{cc}
1\\
z
\end{array}\right),
\end{align*}
satisfies $H\ket{z}=r\ket{z}$. We can then write
\begin{align*}
U=\frac{1}{(1+|z|^2)^{1/2}}\left(\begin{array}{cc}
1 & -\bar{z}\\
z & 1
\end{array}\right),
\end{align*}
clearly,
\begin{align*}
(U^{-1}dU)_{21}=\frac{dz}{(1+|z|^2)},
\end{align*}
and
\begin{align*}
|(U^{-1}dU)_{21}|^2=\frac{dzd\bar{z}}{(1+|z|^2)^2}=\frac{1}{4}d\Omega^2,
\end{align*}
where $d\Omega^2=\delta_{\mu\nu}dn^{\mu}dn^{\nu}$ is the round metric on the sphere. Denoting $n\equiv n_1=1/(e^{-r}+1)$ we have $n_2=1-n_1$, so,
\begin{align*}
ds^2=\frac{1}{4}\tanh^2(r)\big[n^2+(1-n)^2\big]d\Omega^2+\frac{1}{2}n(1-n)dr^2.
\end{align*}
Also  
\begin{align*}
n(1-n)=\frac{1}{(e^{-r}+1)(e^r+1)}=\frac{1}{2\big(\cosh(r)+1\big)},
\end{align*}
and
\begin{align*}
n^2+(1-n)^2 &=\frac{(e^{r}+1)^2+(e^{-r}+1)^2}{(e^{r}+1)^2(e^{-r}+1)^2}=\frac{1}{2}\frac{\cosh(r)}{\cosh^2\big(\frac{r}{2}\big)}=\frac{\cosh(r)}{1+\cosh(r)}
\end{align*}
so
\begin{align*}
\frac{1}{4}\tanh^2(r)\big[n^2+(1-n)^2\big]d\Omega^2&=\frac{1}{4}\frac{\cosh(r)}{1+\cosh(r)}\tanh^2(r)d\Omega^2\\
&=\frac{1}{4}\frac{\sinh^2(r)}{(1+\cosh(r))\cosh(r)}d\Omega^2\\
&=\frac{1}{4}\frac{\cosh(r)-1}{\cosh(r)}d\Omega^2
\end{align*}
Therefore,
\begin{align*}
ds^2=\frac{1}{4}\Big(\frac{\cosh(r)-1}{\cosh(r)}d\Omega^2+\frac{dr^2}{\cosh(r)+1}\Big).
\end{align*}
\subsection{Thermal states}
With the replacement $x^{\mu}\sigma_{\mu}\to \beta E n^{\mu}\sigma_{\mu}$, we get
\begin{align*}
ds^2&=\frac{1}{4}\Big(\frac{\cosh(\beta E)-1}{\cosh(\beta E)}d\Omega^2+\frac{d(\beta E)^2}{\cosh(\beta E)+1}\Big)\\
&=\frac{1}{4}\Big(\frac{\cosh(\beta E)-1}{\cosh(\beta E)}d\Omega^2+\frac{\beta^2 d E^2}{\cosh(\beta E)+1}+2\frac{d\beta dE}{\cosh(\beta E)+1}+\frac{E^2d\beta^2}{\cosh(\beta E)+1}\Big)
\end{align*}
Notice that since $ds^2=\langle \big(\psi^{\dagger} K\psi\big)^2\rangle -\langle \psi^{\dagger} K\psi\rangle^2$, we have that a variation along $\beta$ produces,
\begin{align*}
\psi^{\dagger}K\psi=-\frac{1}{2}d\beta \psi^{\dagger}H\psi,
\end{align*}
and hence 
\begin{align*}
&\langle \big(\psi^{\dagger} K\psi\big)^2\rangle -\langle \psi^{\dagger} K\psi\rangle^2=\frac{1}{4}d\beta^2\big(\langle (\psi^{\dagger} H \psi)^2\rangle -\langle \psi^{\dagger} H\psi \rangle\big)=\frac{1}{4}d\beta^2 \frac{\partial ^2\log Z}{\partial \beta^2}\\
&=-\frac{1}{4}d\beta^2 \frac{\partial \langle \psi^{\dagger}H\psi \rangle }{\partial \beta} =-d\beta^2\frac{1}{4}\frac{\partial T}{\partial \beta } \frac{\partial E}{\partial T}\\
&=d\beta^2 \frac{1}{4}T^2\frac{\partial E}{\partial T}=d\beta^2 \frac{T^2}{4}c,
\end{align*}
where $c=\partial E/\partial T$ is the specific heat of the system.\\
\indent Now if we have a Chern insulator in $2$D as in the main text, we will have, $\mathcal{H}=\int_{\text{B.Z.}}d^2k/(2\pi)^2\psi^{\dagger}_{\bf{k}}H(\bf{k})\psi_{\bf{k}}$ with $H(\bf{k})=d^{\mu}(\bf{k})\sigma_{\mu}=E(\bf{k})n^{\mu}(\bf{k})\sigma_{\mu}$ depending on some parameters $\{t_i\}$, and the Bures metric will be 
\begin{align*}
ds^2=&\int_{\text{B.Z.}}\frac{d^2k}{(2\pi)^2} \Big[\frac{1}{4}\Big(\frac{\cosh(\beta E(\bf{k}))-1}{\cosh(\beta E(\bf{k}))}\delta_{\mu\nu}\frac{\partial n^{\mu}(\bf{k})}{\partial t_{i}}\frac{\partial n^{\mu}(\bf{k})}{\partial t_{j}} dt_{i}dt_{j}\\
&+\frac{E^2(\bf{k})d\beta^2}{\cosh\big(\beta E(\bf{k})\big)+1}+2\frac{\beta E(\bf{k}) \frac{\partial E (\bf{k})}{\partial t_i}dt_id\beta}{\cosh\big(\beta E(\bf{k})\big)+1}+\frac{\beta^2\frac{\partial E(\bf{k})}{\partial t_i}\frac{\partial E(\bf{k})}{\partial t_j} dt_idt_j}{\cosh\big(\beta E(\bf{k})\big)+1}\Big)\Big].
\end{align*}
We remark that for the case of the topological superconductor considered in the main text the charge is only conserved modulo $2$. However, since the expression for the fidelity is the same (as a function of $\vec{d}$ and $\beta$) as the charge conserving ones, the resulting susceptibility has the same functional form. Related to this derivation, we refer the reader to Ref.~\cite{car:spa:val:18} to a recent derivation of the symmetric logarithmic derivative of fermionic Gaussian states by Carollo \textit{et al.}.

\bibliographystyle{unsrt}
\bibliography{TopPhaseTrans_V0}

\end{document}